\documentclass[aps, floatfix, onecolumn,superscriptaddress,nofootinbib,12pt]{revtex4-1}
\usepackage{amsmath}
\usepackage{hyperref}
\usepackage{amsfonts}
\usepackage{amssymb}
\usepackage{graphicx}
\usepackage{latexsym}
\usepackage{color}
\usepackage{booktabs}
\usepackage{dcolumn}
\usepackage{epsfig}
\usepackage{subfigure}
\usepackage{float}
\usepackage{multirow}
\usepackage{ulem}
\usepackage{xcolor}
\usepackage{epstopdf}
\DeclareMathOperator{\arctanh}{arctanh}
\def\be#1\ee{\begin{align}#1\end{align}} 
\def\be{\begin{eqnarray}}
\def\ee{\end{eqnarray}}
\def\l{\label}

\begin{document}

\title{Dynamical evolution of non-minimally coupled scalar field in spherically symmetric de Sitter spacetimes}

 \author{R. D. B. Fontana}\email{rodrigo.fontana@uffs.edu.br}
 \affiliation{Universidade Federal da Fronteira Sul, Campus Chapec\'o, CEP 89802-112, Chapec\'o, SC, Brazil}
 \author{Jeferson de Oliveira}\email{jeferson@gravitacao.org}
 \affiliation{ Instituto de F\'isica, Universidade Federal de Mato Grosso, CEP 78060-900, Cuiab\'a, MT, Brazil}
 \author{A. B. Pavan}\email{alan@unifei.edu.br}
 \affiliation{Universidade Federal de Itajub\'a, Instituto de F\'isica e Qu\'{\i}mica, CEP 37500-903, Itajub\'a, MG, Brazil}

\date{\today}
\begin{abstract}

We investigate the dynamical behavior of a scalar field non-minimally coupled to Einstein's tensor and Ricci scalar in geometries of asymptotically de Sitter spacetimes. We show that the quasinormal modes remain unaffected if the scalar field is massless and the black hole is electrically chargeless. In the massive case, the coupling of both parameters produces a region of instability in the spacetime determined by the geometry and field parameters. In the Schwarzschild case, every solution for the equations of motion with $\ell>0$ has a range of values of the coupling constant that produces unstable modes. The case $\ell=0$ is the most unstable one, with a threshold value for stability in the coupling. For the charged black hole, the existence of a range of instability in $\eta$ is strongly related to geometry parameters presenting a region of stability independent of the chosen parameter.
\end{abstract}
\maketitle


\section{Introduction}\label{intro}

The evolution of probe fields in black hole backgrounds has long been a very active field of research in theoretical physics \cite[and references therein]{Berti:2009kk,Kokkotas:1999bd,Nollert:1999ji}. Probe field profiles in the time domain present a discrete set of complex frequencies called quasinormal frequencies (QNFs) that can provide valuable information about the structure of spacetime. Each of these frequencies corresponds to a damped vibrational mode of the field, the so-called quasinormal mode (QNM). The set of QNM's carry specific information about the signature of the geometry (e. g. black hole solutions) and its interaction with fields, since it depends on the parameters that define the metric.

The applications of quasinormal modes are manifold: probing the linear stability of black holes and stars spacetimes \cite{esta}; identification of astrophysical black holes through gravitational waves signals \cite{Nollert:1999ji}, experimentally verified by LIGO \cite{Abbott:2016blz}  \cite{TheLIGOScientific:2017qsa}; studying the role played by such oscillations in the context of gauge/gravity duality, especially in the AdS/CFT \cite{Nunez:2003eq} \cite{Horowitz:1999jd} \cite{Son:2002sd} \cite{kaielciojef} \cite{Sybesma:2015oha} and dS/CFT correspondences \cite{elciobinlimaqiu,elciokarluciolima}.

The stability of black holes and stars has been discussed in several works \cite{chandrasekhar} since the 50's with the original paper of Regge and Wheeler analyzing the Schwarzschild singularity \cite{Regge:1957td}. The QNM's of scalar, Abelian gauge, and fermionic free probe fields evolving in the neighborhood of black holes have also been used to obtain insights about the nature of spacetime. In the case of asymptotically flat black holes these QNM's are, by the no-hair theorems, functions of only the mass $M$, the electric charge $Q$, and the angular momentum $L_\phi$ of the black hole \cite{Bekenstein:1996pn}. However, more recently, these theorems were circumvented in the asymptotically AdS black holes and other configurations with non-minimally coupled fields such that hairy black holes solutions have been found \cite{Gubser:2008px,Abdalla:2010nq,Rinaldi:2012vy,Minamitsuji:2013ura,Volkov:1998cc}. In the latter cases, the QNM's depend on other hairs of the spacetime, and black hole phase transitions are present.

In AdS/CFT correspondence context, a robust interpretation for the QNM spectra in the view of a quantum field theory at finite temperature (defined at the AdS boundary) is provided: the inverse of the imaginary part of the fundamental quasinormal frequency is understood as a relaxation time of the dual operator at the boundary \cite{Horowitz:1999jd}. Among the applications of AdS/CFT correspondence to condensed matter physics \cite[and references therein]{Hartnoll:2009sz}, we mention the phase transitions at the border theory giving rise to the so-called holographic superconductors \cite{Hartnoll:2008kx,Hartnoll:2008vx,Lin:2014bya,Abdalla:2013zra,Pan:2009xa,holo}: the phenomena yields a specific bulk effect through the QNMs, i.e, growing/decaying oscillations of a given probe field in the bulk correspond to a conductive/superconducting phase at the dual field theory \cite{Abdalla:2010nq}. The presence of instabilities (growing modes) in the quasinormal spectrum therefore indicates a phase transition at the border.

On the dS/CFT correspondence \cite{strominger}, the evolution of probe fields on the gravity side is related to fundamental quantities in the border field theory \cite{elciokarluciolima} \cite{elciobinlimaqiu}: the poles of the two-point correlator of the three-dimensional conformal field theory at the boundary scale perfectly the QNMs spectrum of a massive scalar field in the de Sitter spacetime.

Non-minimally coupled (NMC) curvature models were firstly considered in the late 80's \cite{amendoa1}, as an alternative gravitation theory. The presence of a scalar field coupled to curvature terms in Einstein-Hilbert action allows for a suitable solution for the inflation exit, and in general has a de Sitter spacetime as the attractor for later times, as should be expected. Besides the traditional terms of NMC models, a few years later, derivative terms were introduced in the action \cite{amendoa2}, expanding the possibilities for the scalar field potential, characterizing the non-minimally derivative coupling (NMDC) models. From the possible derivative terms, only two significant contributions are in general considered. With a particular scale of the Lagrangian couplings and the cosmological constant, the inflation scenario is generated, as well as the de Sitter spacetime remnant from the curvature  equations \cite{capozzielo,capozzielo2}.

The curvature equations coming from NMDC models are of third or higher order, in general. For a particular choice of couplings, however, it is still possible to achieve second order equations: when the Lagrangian derivative terms are placed as Einstein tensor coupled to scalar field components \cite{sushkov}. This choice turns the NMDC into a more suitable (simple) form, as it makes unnecessary to fine tune the scalar field potential.

Beyond the strategic elimination of the fine tuning problem, another possible purpose of the coupling is to perform as a dark matter component, feasible in the form of $\Lambda$CDM model \cite{gao}. The rate of the scalar field density and total density in the model is slightly different from that of a cold dark matter model, but still in the observationally allowed range. Once NMDC models could be used to describe dark energy and dark matter, they would be instrumental to understand how this coupling affects black holes: for instance, in the context of scalar-tensor gravity exact hairy black hole solutions have been found using NMDC models \cite{Rinaldi:2012vy,Minamitsuji:2013ura}.

In the case of NMDC models, field propagation and quasinormal modes were investigated in a group of papers with a different approach \cite{quasi1,quasi2,quasi3}. In \cite{quasi1,quasi2} the QNMs were obtained in spacetimes with charge, mass, dilaton fields and other hairy geometries. In \cite{Chen, quasi3} the dynamical evolution of scalar and vector fields are examined showing the presence of dynamical instabilities associated with a critical value of the NMDC coupling.

In this work we concern ourselves with the dynamical evolution of a scalar field in different geometries with the non-minimal derivative coupling introduced in the action as
\begin{equation}
\label{s2e1}
S = \int d^4 x \sqrt{-g}\bigg( F(\Phi ,\mathcal{R}, \mathcal{R}_{\mu \nu} \mathcal{R}^{\mu \nu},\mathcal{R}_{\mu \nu \delta \sigma}\mathcal{R}^{\mu \nu \delta \sigma}) + H( \Phi , \partial_\sigma \Phi \partial^\sigma \Phi , \nabla^2 \Phi ) + V(\Phi )\bigg),
\end{equation}
where the function $F$ encodes all possible Lagrangian curvature terms along with their couplings to a scalar field component $\Phi$,  $ H$ gives a general coupling between curvature and the scalar field kinetic term, and $V(\Phi)$ is the scalar field potential.
This action is relevant, e.~g., in the context of quantum gravity \cite{Buchbinder:1992rb}, where additional terms in the curvature of  most second degree are added to the  Einstein-Hilbert Lagrangian. Although the correspondent gravity theory is not unitary, it can be considered as the starting point of an effective theory of gravity, since the description of it does not have to satisfy all requirements imposed by the fundamental physics.
As a particular case of action (\ref{s2e1}), we consider the simplest NMDC model with matter terms as follows:
\begin{equation}
\label{s2e2}
S=\int d^4 x \sqrt{-g}\bigg({\cal{L}}_\text{background}(\mathcal{R},\Lambda,F^{\mu\nu})+{\cal{L}}_\text{perturbative}(\Phi)\bigg) ,
\end{equation}
where
\begin{equation}
\label{s2e3}
{\cal{L}}_\text{background}(\mathcal{R},\Lambda,F^{\mu\nu})=\frac{\mathcal{R}}{16\pi G} - \frac 6 {L^2} -\frac{F_{\mu\nu}F^{\mu\nu}}{4},
\end{equation}
with $\mathcal{R}$ standing for the Ricci scalar, $L$ is the dS radius and $F_{{\mu\nu}}$ are the components of electromagnetic field strength tensor. Also,
\begin{equation}
\label{s2e4}
{\cal{L}}_\text{perturbative}(\Phi)= -\frac{1}{2}\left(g^{\mu\nu}+\eta G^{\mu\nu}\right)\partial_{\mu}\Phi\partial_{\nu}\Phi-\frac{1}{2}\mu^2\Phi^2-V_{int}(\Phi ),
\end{equation}
where $g_{\mu\nu}$ and $G_{\mu\nu}$ are, respectively, the components of metric and Einstein tensors, $\Phi$ is the probe scalar field with mass $\mu$ and $\eta$ is the NMDC parameter.

Here we are interested in the effect produced on the scalar field equation, given usual black hole geometries as a fixed background. In this approach the probe fields are treated as small perturbations, that are not expected to change the fixed geometry and decay in time. In such case, the corrections of the metric elements are of small order and can be consistently set to zero \cite{Berti:2009kk}, once the energy-momentum tensor for the scalar field is quadratic.

The paper is organized as follows: in section \ref{sec1} we establish a general equation of motion for the scalar field $\Phi$ for spherically symmetric spacetimes. In sections \ref{sec2}-\ref{sec4} we analyze the dynamical properties of the field in the spacetimes of de Sitter, Schwarzschild-de Sitter, and Reissner-Nordstr\"om-de Sitter. In section \ref{sec6} we present our conclusions and final remarks relative to peculiar features of the coupling for all geometries considered.


\section{Equation of motion}\label{sec1}

We first consider the four-dimensional spherically symmetric black hole (or de Sitter) background solution, namely
\begin{align}
\label{metric}
ds^2 = -f\ dt^2+\frac{1}{f}\ dr^2+r^2\ d\Omega^2,
\end{align}
with $d\Omega^2=d\theta^{2}+\sin^2\theta d\phi^2$ representing the 2-sphere line element and $f = f(r)$. The equation of motion for the scalar field $\Phi$ derived from the action \eqref{s2e2} is given by
\begin{align}
\label{eqm}
\frac{1}{\sqrt{-g}}\partial_{\mu}\left(\frac{}{}\sqrt{-g}\ h^{\mu\nu}\ \partial_{\nu}\Phi\right)-\frac{d\tilde{V}}{d\Phi} =0,
\end{align}
where the potential of the scalar field is given by $\tilde{V}=\frac{1}{2}m^2\Phi^2$, and we redefine the mass term as $m^2\rightarrow \mu^2+\xi\mathcal{R}$, being the last term originated by $V_{int}$ in (\ref{s2e4}). We also have
\begin{align}
\label{eqm2}
h^{\mu\nu}=g^{\mu\nu}+\eta\ G^{\mu\nu},
\end{align}
which acts as an induced metric for the scalar field equation, through $\eta$. The convention of sign used for $\eta G^{\mu \nu}$ along this work is the same of that used in reference \cite{sushkov} ($\eta \rightarrow \kappa$) and contrary to \cite{quasi1,quasi3}, ($\eta \rightarrow -\beta$). From the cosmological point of view, both scenarios are explored in \cite{sushkov},  $\eta > 0$ and $\eta < 0$. We studied cases with $\eta >0$ (correspondingly $\kappa >0$) where the Universe has a quasi-de Sitter behavior gracefully solving the problem of exit of inflation era. 

Applying the standard Ansatz to separate variables in spherically symmetric spacetimes we write the field in radial-temporal and angular parts,
\begin{align}
\label{eqm3}
\Phi(t,r,\theta,\phi)=\sum_{\ell , m_\phi } R(r,t)\ Y_{\ell , m_\phi}(\theta ,\phi ),
\end{align}
which, introduced into Eq.~\eqref{eqm}, yields
\begin{align}
\label{eqm4}
-\frac{\partial^2 R}{\partial t^2}  + \alpha \frac{\partial^2 R}{\partial r^2}  + \alpha \left(\frac{2}{r}+\frac{dF}{dr}\right)\frac{\partial R}{\partial r}- \vartheta (r)R=0,
\end{align}
with the potential $\vartheta (r)$ being
\begin{align}
\label{eqm5}
\vartheta (r)=\frac{\beta\ \ell(\ell+1)}{r^2}+\gamma m^2,
\end{align}
and the functions $\alpha$, $\beta$, $\gamma$ and $F$ given by,
\begin{align}
\label{eqm6}
\alpha&=f^2,\\
\beta&=\frac{1+\eta B}{1-\eta A}f,\\
\gamma&=\frac{f}{1-\eta A},\\
F&=\ln \big( (1-\eta A)f \big).
\end{align}
Functions $A$ and $B$ are specific of each geometry and defined in the appendix. In order to place~\eqref{eqm4} as a Schr\"odinger-like equation, we perform a change in the radial coordinate to the tortoise system, $\dfrac{dr_{*}}{dr}=\dfrac{1}{f}$ resulting in
\begin{align}
\label{eqmstar}
-\frac{\partial^2 \tilde{R}}{\partial t^2}  + \frac{\partial^2 \tilde{R}}{\partial r_{*}^2} + V(r)\tilde{R}=0,
\end{align}
where $R=\dfrac{\tilde{R}(r,t)}{r\sqrt{k}}$ and $k=1-\eta A$. The tortoise coordinate system has the advantage of avoiding singularities in the integration of the scalar field equation encapsulating it beyond the event horizon. In this case, the above effective potential is written as
\begin{align}
\label{effecetivepotstar}
V(r) &= \frac{\alpha}{16} \left(\frac{4 \eta ^2 A'^2}{(-1+\eta  A)^2}-\frac{16 f'}{r f}+\frac{4 \eta  A' \left(4 f+ r f'\right)}{r (1-\eta  A) f}\right.\nonumber \\
&-\left. \frac{2 \eta  A' \left(-\eta  f A'+2 (1-\eta  A) f'\right)}{(1-\eta  A)^2 f}+\frac{8 \eta  A''}{1-\eta  A}\right)- \vartheta (r),
\end{align}
which allows us to integrate and use different methods to attain the scalar field profiles in the time domain as well as the quasinormal spectra.

\section{QNM's for non-minimally coupled scalar fields evolving in the pure de Sitter and Schwarzschild-de Sitter spacetimes}\label{sec2}
In this section, we are going to explore the dynamics of non-minimally coupled scalar field in de Sitter and Schwarzschild-de Sitter spacetimes, through the computation of quasinormal frequencies spectrum and modes.

\subsection{De Sitter spacetime}

We firstly analyze the pure dS case, in which an analytical expression for the scalar QNMs was found in \cite{Bin}, where the probe scalar field is not coupled to the Einstein tensor ($\eta = 0$ in the Lagrangian~\eqref{s2e4}).

In ($3+1$) dimensions, the line element of dS spacetime can be cast as
\begin{align}
\label{metricdS}
ds^2 = -\left(1-\frac{r^2}{L^2}\right)\ dt^2+\frac{1}{\left(1-r^2/L^2\right)}\ dr^2+r^2\ d\Omega^2,
\end{align}
where $L$ stands for the dS radius, related to the cosmological constant $\Lambda$ by $L^2=3/\Lambda$. Considering the evolution of a probe scalar field with mass $m$ in the pure dS geometry, the corresponding effective potential reads as
\begin{align}
V(r)= \frac{\mathcal{C}_{1}}{L^2 \cosh^2(r_{*}/L)}+\frac{\mathcal{C}_{2}}{L^2 \sinh^2(r_{*}/L)},
\end{align}
with the radial tortoise coordinate $r_{*}=L \arctanh{\left(r/L\right)}$, $\mathcal{C}_{1}=-2+m^2L^2$, and
$\mathcal{C}_{2}=\ell(\ell+1)$.

We can generalize the results for the scalar QNFs found in \cite{Bin}, $\omega_{I,II}$, considering an NMDC term
$\eta \neq 0$:
\begin{align}
\nonumber
\omega_{I}&=-\frac{i}{L}\big( 2n+\ell+h_{\pm} \big) ,\\
\label{pureds1}
\omega_{II}&=-\frac{i}{L}\big( 2n-(\ell+1)+h_{\pm}\big) ,
\end{align}
where,
\begin{align}
\label{pureds2}
h_{\pm}&=\frac{3}{2}\pm\sqrt{\frac{9}{4}-\frac{m^{2}L^{4}}{L^{2}-3\eta}}.
\end{align}

From these exact expressions it is clear that the behavior of the coupling parameter $\eta$ will affect the quasinormal spectrum. If $\eta$ is bounded by $\eta< L^{2}/3$, the range of allowed values for $m$ in order to have QNFs with non-null real part is
\begin{align}
\label{pureds3}
m>\frac{3}{2L}\sqrt{1-\frac{ 3\eta}{L^{2}}},
\end{align}
which constraints the field mass to be positive definite. Using the expression~\eqref{pureds2} into the
Eqs~\eqref{pureds1}, the two sets of QNFs can be cast in the form
\begin{align}
\label{pureds4}
\omega_{I}=\pm\frac{1}{L}\left(\frac{m^2L^4}{L^2-3\eta}-\frac{9}{4}\right)^{1/2}-\frac{i}{L}\left(2n+\ell+\frac{3}{2}\right),\\
\label{pureds5}
\omega_{II}=\pm\frac{1}{L}\left(\frac{m^2L^4}{L^2-3\eta}-\frac{9}{4}\right)^{1/2}-\frac{i}{L}\left(2n-\ell+\frac{1}{2}\right).
\end{align}
These expressions generalize the previous results found for the scalar field trivially coupled to the geometry \cite{Bin}. In what follows, we show the existence of a region of parameters in which purely imaginary and unstable QNMs arise in the system, the origin of the instabilities is attributed  to the non-canonical coupling between the scalar field and the dS geometry.

\subsubsection{Purely imaginary frequencies and instabilities}

Using the expressions for the frequencies found above, we constraint the values of the NMDC parameter $\eta$ and the scalar field mass $m$ in order to get purely imaginary QNMs and, more interesting, a range of parameters allowing growing modes, i.e., frequencies with positive imaginary part.

Purely imaginary frequencies have been found in the context of black hole perturbations, and its applications to the AdS/CFT correspondence are manifold. In \cite{kdv} the authors found a close relation between the Korteweng-de Vries equation and the three dimensional Lifshitz black hole in New Massive Gravity (NMG). They also showed that the scalar QNMs in the hydrodynamic limit are purely imaginary, which in the view of linear response theory corresponds to a solitonic solution. Also in the context of NMG, purely imaginary QNMs were found beyond the hydrodynamic limit in \cite{berjefpell}. Furthermore, purely imaginary spectra have been found for a probe scalar field evolving on the geometry of $d$-dimensional Lifshitz black hole \cite{owenfideljef} and for the Chern-Simmons sector of $d$-dimensional Lovelock black holes \cite{kaielciojef}.

An attempt to give an interpretation of QNMs in the framework of the dS/CFT correspondence \cite{strominger} was made in \cite{elciobinlimaqiu}, where the authors considered the exact QNM spectrum of scalar perturbations on a three-dimensional rotating dS black hole  and in \cite{elciokarluciolima} for a pure $d-$dimensional dS black hole. In \cite{elciobinlimaqiu}, it was found an exact relation between the QNM spectrum and the spectrum of thermal excitations of a Conformal Field Theory, which presents growing modes, leading to regions of instability.
Following the same procedure as in \cite{elciobinlimaqiu}, it is possible to show that there are growing modes and regions of instability in the case of the $4-$dimensional dS spacetime with $\eta\neq 0$.

If we take $L=1$ and $\ell=0$ in the first set of QNMs $\omega_{I}$ \eqref{pureds4}, the condition to obtain purely imaginary QNFs is that $\eta> 1/3$, thus
\begin{equation}\label{freq_1p}
\omega_{I}=i\left( \pm\left(\frac{m^{2}}{3\eta-1}+\frac{9}{4}\right)^{1/2} -\left(2n+\frac{3}{2}\right) \right).
\end{equation}
Considering then the fundamental mode $n=0$ for the positive branch of $\omega_{I}$ and setting for simplicity $\eta=2/3$, we find that the fundamental QNF corresponds to a purely growing mode for $m^2>0$. For $\eta\rightarrow \infty$, the QNFs of the positive branch is bounded by $\omega_{I}^{+}=-2n i$, while in the negative branch $\omega_{I}^{-}$ we have only decaying QNFs for positive masses, bounded by $\omega_{I}^{-}=-(3+2n)i$.

The same analysis can be done for the second set of QNFs \eqref{pureds5}, leading to
\begin{equation}
\label{freq_2p}
\omega_{II}=i\left( \pm\left(\frac{m^{2}}{3\eta-1}+\frac{9}{4}\right)^{1/2} -\left(2n+\frac{1}{2}\right) \right).
\end{equation}

For the positive branch, the fundamental QNF is a growing mode for $m^2>0$ (setting $\eta>2/3$) and for the negative branch there is only QNFs with negative imaginary part. When $\eta\rightarrow \infty$, these frequencies are bounded by $\omega_{II}^{-}=-(2n+2)i$ and $\omega_{II}^{+}=-(2n-1)i$.

In short, growing purely imaginary QNMs in the positive branch of the two sets of exact frequencies are present in the spectra, featuring two regions of instability. 

The result for the poles of the two-point correlation function in $4$- dimensional dS spacetime found in \cite{elciokarluciolima} can be easily generalized for the case of a non-vanishing parameter $\eta$ by means of the following rescaling of the scalar field mass,
\begin{equation}
 m^{2}\rightarrow\frac{m^{2}L^{2}}{L^{2}-3\eta},
\end{equation}
in which case the poles are written as
\begin{equation}
\label{poles1}
\omega=\pm\frac{i}{L}\left(h_{\pm}+ \ell +2n\right),
\end{equation}
where $h_{\pm}$ is given in \eqref{pureds2}. The above expression is equivalent to the first set of QNMs  \eqref{pureds1}, therefore, the positive branch of the poles coincides with the region of instability discussed in the preceding analysis. Such a result seems to be in agreement with the dS/CFT correspondence, namely, the regions of instability in the bulk QNM spectrum matches with those obtained by the calculation of Hadamard two-point function.

%
%
%
%
%
%

\subsection{Schwarzschild-de Sitter spacetime}\label{sec3}

The coupled-scalar field equation introduced in section \ref{sec1}, can also be studied in a Schwarzschild-de Sitter geometry (SdS), being $f(r)= 1-(2M/r)-(r^2/L^2)$ the $g_{tt}$ element of the metric. In such case, the extra functions in the Klein-Gordon equation read
\begin{align}
\label{sds4}
\alpha=f^2, \qquad \beta=f,\qquad \gamma=\frac{f}{\left(1-\frac{3\eta}{L^2}\right)},\qquad F=\ln\left( \left(1+\frac{3\eta}{L^2}\right)f\right),
\end{align}
and the potential of equation (\ref{eqm4}) reads
\begin{align}
\label{sds6}
\vartheta (r)=f(r)\ \left( \frac{\ell(\ell+1)}{r^2}+\frac{\mu^2 L^2+12\xi}{L^2-3\eta}\right).
\end{align}
In the Schwarzschild-de Sitter case, the field transformation as introduced in section \ref{sec1} is given by $R\rightarrow \frac{\tilde{R}}{r\sqrt{k}}$, and, as $k$ is constant, it may be ignored in the scalar field equation. Then the wave equation is the same as (\ref{eqmstar}) with the effective potential given as
\begin{align}
\label{sds9}
V(r)=\left(1-\frac{2M}{r}-\frac{r^2}{L^2}\right)\ \left( \frac{\ell(\ell+1)}{r^2}+\frac{m^2 L^2}{L^2-3\eta}+\frac{2M}{r^3}-\frac{2}{L^2}\right) .
\end{align}
Here the $m^2$ term was rescaled as the effective mass of the scalar field, being a function of the ordinary mass $\mu$ and of the Ricci-coupling introduced in the perturbed potential: $m^2=(\mu^2+12\xi/L^2)$. This term is essential in the Schwarzschild case, without which there would be no influence coming from the NMDC term $\eta$ on the equation of motion for the scalar field (a different situation is seen in the Reissner-Nordstr\"om geometry). In the expression (\ref{sds6}), we may still realize that the term $\frac{m^2L^2}{L^2-3\eta}$ act again as the new scalar field mass, becoming positive/negative depending on the parameters of the geometry and field. This fact changes the signal of the effective potential between horizons, what can naturally produces instabilities for the field evolution.

Though in first principle, the instability of the spacetime to the scalar field perturbation is not dependent on the multipole number - in the sense that the presence of only one multipole turns the field unstable - numerically, this is not the case. For highly enough $\eta$ and $\ell > 0$, the field turns out to be stable no matter the geometry parameters. This is not the case however for $\ell = 0$, as we may further discuss.

As discussed all along in the literature, the usual evolution of the scalar field after a initial burst in a positive potential is that of a damped oscillator, what characterizes the quasinormal modes. In the pure Schwarzschild-de Sitter case the massive scalar field chooses one of the three different behaviors after the ringing phase: (1) decays exponentially ($\ell > 0$), (2) goes to a constant value that scales the cosmological constant ($\ell = 0$), (3) oscillates indefinitely as a function of the scalar field mass.

In a more general case, however, a different behavior arises when the potential is not entirely positive between horizons: unstable modes can emerge and the geometry is then expected to change. This is the case for the NMDC $\eta$ in Schwarzschild-de Sitter geometry we study here: the potential is partly or entirely negative (depending on the coupling and geometry parameters). In this section, we evolve the field for different $L,\eta$ and $\ell$. All studied cases take $L^2 > 27M^2$, which is the causal structure condition for the presence of an encapsulated singularity (by the event horizon) and a cosmological horizon.

\begin{figure}[!ht]
\begin{center}
		\epsfig{file =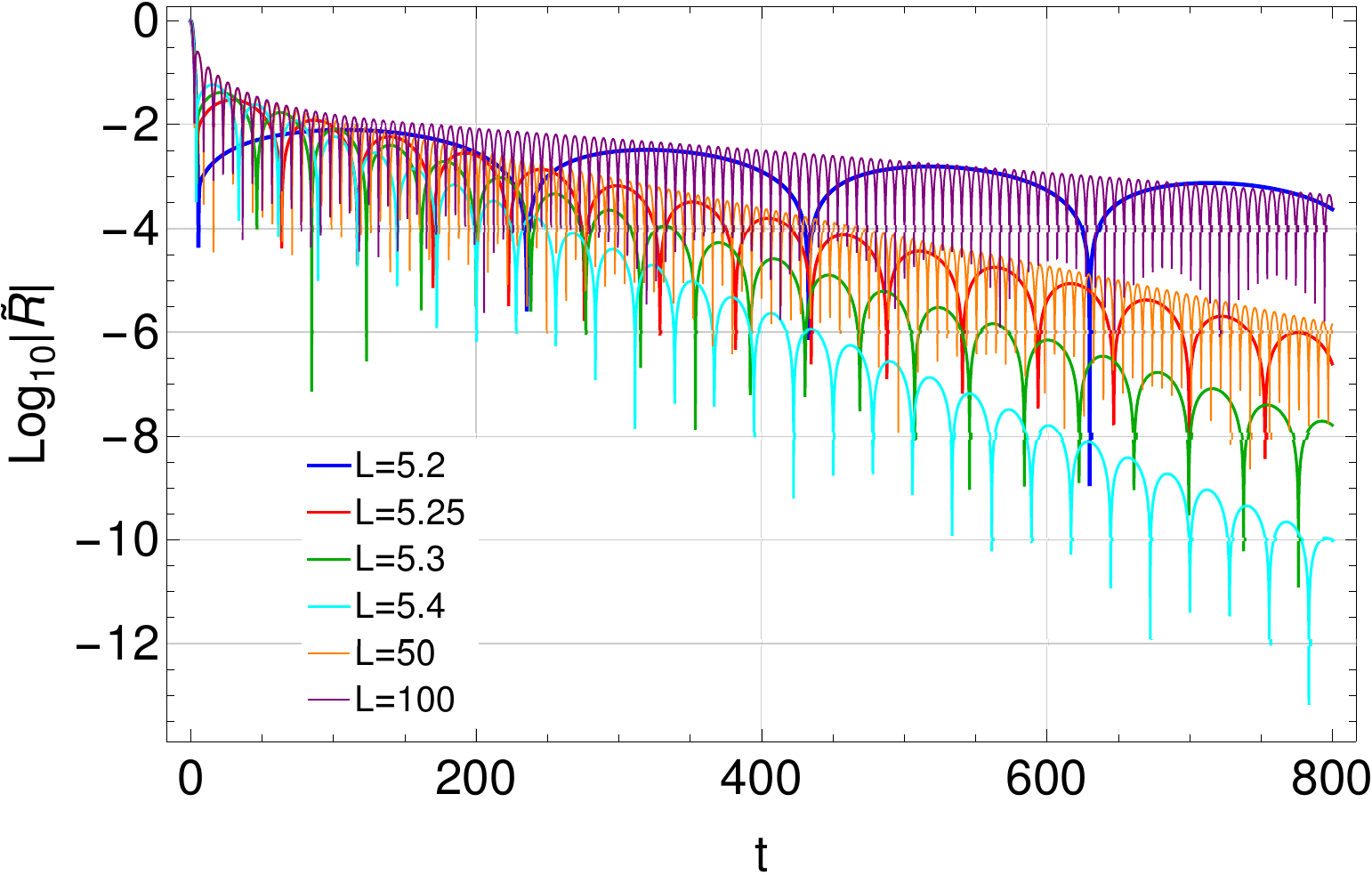, width=6.0cm, height=5.0cm}
		\epsfig{file =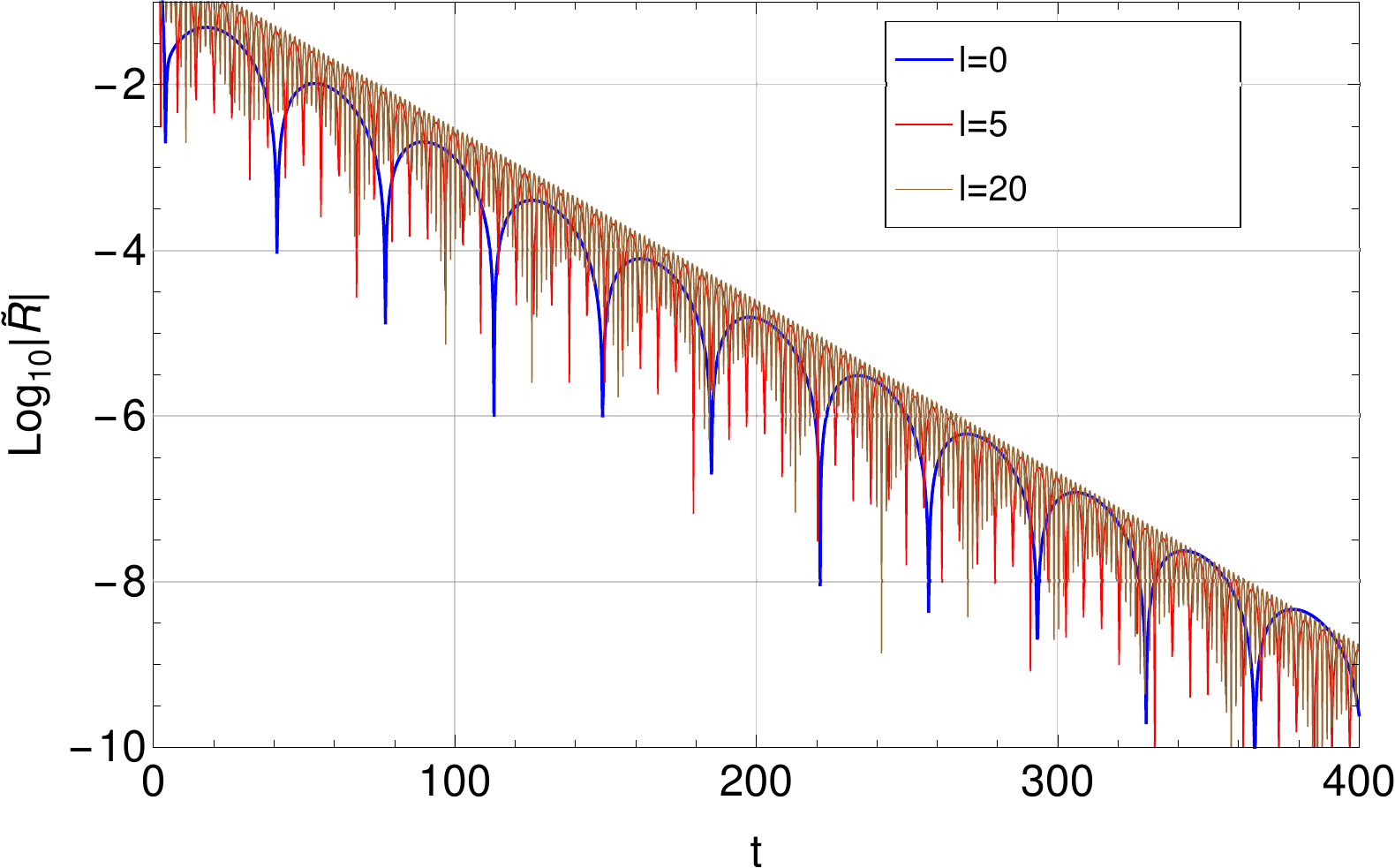, width=6.0cm, height=5.0cm}
\end{center}
\caption{{\small Time evolution of the scalar field with different values of $\ell$ and $L$. The parameters of the geometry are $M= m / 0.3 = \eta = 1$ and $L/6 =1$ (right) $\ell=0$ (left).}}
\label{figsch1}
\end{figure}

In figure \ref{figsch1} (right and left panels) we see typical quasinormal mode evolutions for the scalar field for different values of $\ell$ and $L$: the higher the multipole number/dS radius, the smaller the frequency of oscillation. The imaginary part of $\omega$ varies very slowly with $l$, which is typical for the Schwarzschild-dS geometry also in the absence of couplings, but is majorly affected for the variation of $L$, diminishing as we increase the cosmological radius. The interesting feature is the emergence of an oscillatory evolution, introduced by the NMDC $\eta$ for the $\ell = 0$ mode: there is a quasinormal ringing phase (left panel of figure \ref{figsch1}) which does not exist in the Schwarzschild-dS case \cite{Molina}, associated now entirely with the renormalized mass of the scalar field.

The effect of the cosmological constant is similar in the Schwarzschild-dS case: the higher the $\Lambda$, the smaller the real part of $\omega$. The behavior for $\omega_I$ is more complicated, oscillating in a given scope of $L$ and becoming arbitrarily small as $\Lambda$ increases. This can be seen  in figure \ref{figsch1} (left panel), and in table~\ref{tsch1}, which lists quasi-frequencies for different values of $\ell$ and $L$. In the same table we can also see different $\omega$ for a range of $\eta$: the asymptotic values of the coupling are the same as for the massless scalar field propagation in Schwarzschild-dS case.


\begin{table}[ht]
	\caption{Fundamental quasinormal modes for non-minimally coupled scalar field evolving in Schwarzschild-dS black holes.}
	\label{tsch1}
	\begin{tabular*}{\columnwidth}{*{6}{c@{\extracolsep{\fill}}}}
		\hline
		\multicolumn{2}{c}{$M=\eta = L/6 = m/0.3 = 1$} & \multicolumn{2}{c}{$M=\ell =m/0.5=\eta/2=1$} & \multicolumn{2}{c}{$M=\ell =L/9=m/0.5=1$} \\
		\hline
		\hline
		$\ell$ & $\omega$ & $L$ & $\omega$ & $\eta$ & $\omega$ \\
		\hline
		$0$ &  $9.695-0.04775i$ & $5.2$ & $0.01590-0.003730i$ & $0$ & $0.1982-0.04479i$\\
		$1$ &  $4.909-0.04658i$ & $5.25$ & $0.05927-0.01367i$ & $5$ & $0.2352-0.04397i$\\
		$2$ & $1.977-0.04787i$ & $5.3$ & $0.08177-0.01876i$ & $10$ & $0.5381-0.04349i$\\
		$3$ &  $0.5339-0.04797i$ & $5.4$ & $0.1131-0.02565i$ & $11$ & $0.3907-0.04344i$\\
		$4$ & $0.4386-0.04792i$ & $6$ & $0.2092-0.04446i$ & $50$ & $0.1966-0.05852i$\\
		$5$ &  $0.3440-0.04781i$ & $10$ & $0.3659-0.05512i$ & $55$ & $0.1301-0.05125i$\\
		$20$ &  $0.2506-0.04757i$ & $30$ & $0.4469-0.02382i$ & $56$ & $0.1105-0.5731i$\\
		50 & $0.1608-0.04688i$ & $50$ & $0.4588-0.01281i$ & $500$ & $0.1100-0.05751i$\\
		$100$ &  $0.08713-0.04506i$ & $100$ & $0.4694-0.004889i$ & $5000$ & $0.1318-0.05088i$ \\
		\hline
	\end{tabular*}
\end{table}

\begin{figure}[!ht]
	\begin{center}
		\epsfig{file =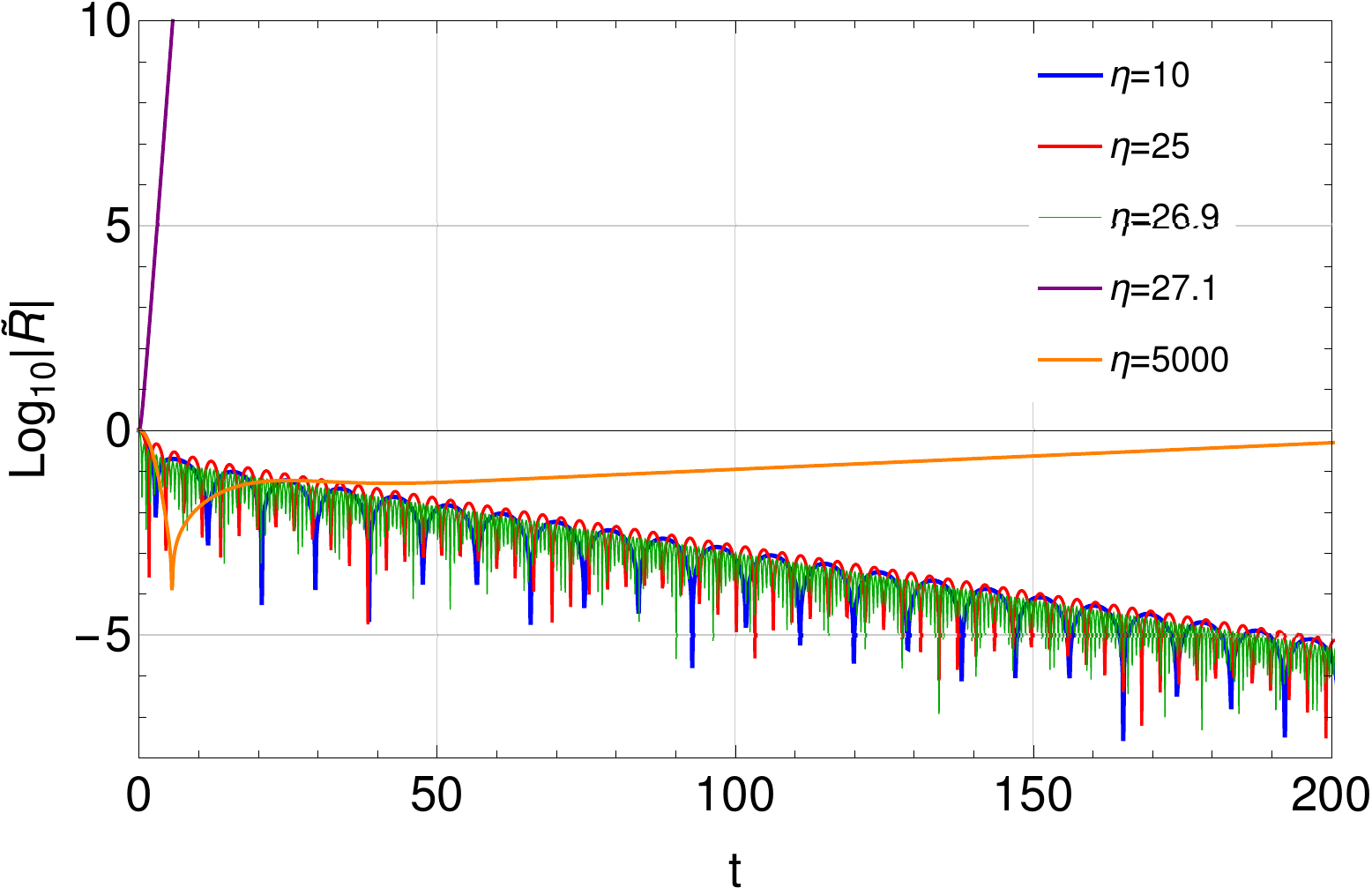, width=6.0cm, height=5.0cm}
		\epsfig{file =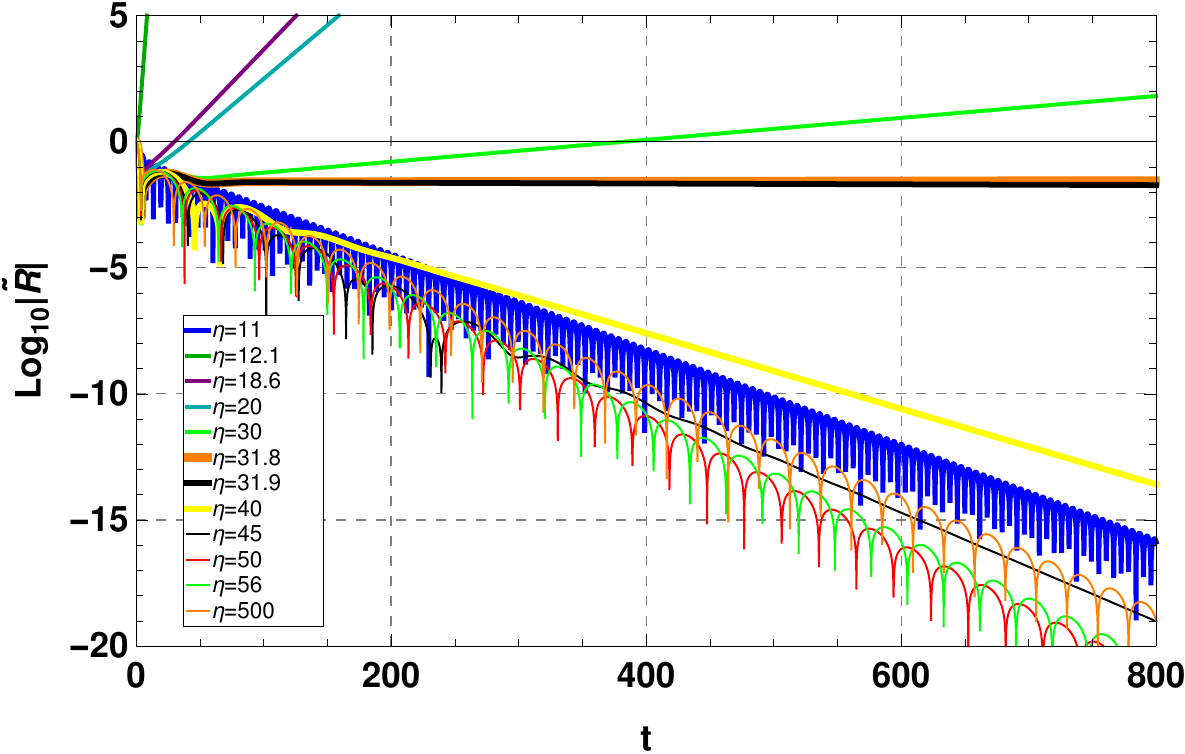, width=6.0cm, height=5.0cm}
	\end{center}
	\caption{{\small Time evolution of the scalar field for different values of $\ell$ and $\eta$. Geometry parameters read $M=\ell = m / 0.5 = L /9 = 1$ (right panel) and $M=L/9 = m / 0.5 = 1$, $\ell=0$ (left panel).}}
	\label{figsch2}
\end{figure}

In the left panel of figure \ref{figsch2} we see the transition between stable/unstable dynamics as a function of $\eta$ for the special case $\ell =0$. Stable evolution takes place from $\eta = 0 $ until $\eta < L^2/3 = 27$, exhibiting the expected decay in time (the potential being only positive). For $\eta > L^2/3$, on the other hand, the dynamics is always unstable: even for asymptotic $\eta$, where the potential is partly positive, there is no stable evolution (see right panel of figure \ref{figsch2}).

The instability comes as no surprise since the effective potential term in such case allows the presence of a negative square mass term: whenever $\eta > L^2/3$, the field becomes unstable.

Although in the Schwarzschild geometry the instability for $\eta > \Lambda^{-1}$ is easily verifiable, the situation changes significantly for $\ell >0$, as it can be seen in the same figure, right panel ($\ell =1$), in which the field evolves unstably for $27< \eta < 31.8$, for the chosen parameters, $M=\ell=L/9=2m=1$, but decays in time for $\eta>31.9$ and the same parameters. Although the fundamental mode destabilizes the geometry from the critical point $\eta = L^2/3$ on, for the excited modes, there is a second critical value present from which the excited modes are stable. The existence of a point for $\eta$ from which the field evolves stably is the same found in the charged black hole as we may, see,  but differently in the Reissner--Nordstr\"om black hole, this fact happens also for $\ell =0$, thus decreasing the region of instability.

The very special case in which $\eta = \Lambda^{-1}$ has no solution different from the trivial one, for the massive Klein-Gordon equation, being identically satisfied in the massless case.

Regarding the transition from stable to unstable evolutions, and further again to stable, this transitional behavior is observed also in the Reissner-Nordstr\"om geometry, namely, the existence of a region of instability for $\eta$. We explore the subject in the next section.
\\

{\it Quasi-extremal regime}
\\

The quasinormal modes for a massive scalar field minimally coupled evolving in Schwarzschild-dS black holes in the quasi-extremal limit (defined by $\delta=\frac{r_c - r_h}{r_h}\ll 1$) have been exactly and numerically calculated by Molina \cite{MolinaII}, Cardoso and Lemos \cite{Cardoso}. We follow their approach and extend the calculation for the NMDC case. The effective potential can be written in terms of the tortoise coordinate $r_{*}$ as
\begin{equation}
\label{potentialextremal}
V(r_{*})=\frac{V_{0}}{\cosh^2(\kappa_{+}r_{*})}
\end{equation}
where $\kappa_+=\frac{1}{2}\frac{df}{dr}|_{r=r_h}$ is the surface gravity at the event horizon $r_h$ and $V_0$ is the saddle point of $V(r)$ given by
\begin{equation}
\label{constantpotentialextremal}
V_{0}=\left(\frac{\ell(\ell+1)}{r_h^{2}}+\frac{m^2L^2}{L^2-3\eta}\right)\frac{(r_{c}-r_h)\kappa_{+}}{2}.
\end{equation}
The effective potential \eqref{potentialextremal} is similar to the P\"oschl-Teller potential for which the QNM's can be exactly obtained with appropriated boundary conditions \cite{MolinaII,Cardoso}, which in our case yields 
\begin{equation}
\label{qnmextremal}
\frac{\omega}{\kappa_{+}}=\sqrt{\left(\frac{\ell(\ell+1)}{r_h^{2}}+\frac{m^2L^2}{L^2-3\eta}\right)\frac{r_{c}-r_h}{2\kappa_{+}}-\frac{1}{4}}- i \left(n+\frac{1}{2}\right).
\end{equation}
The critical behavior of the modes in relation to the non-minimally coupled constant is qualitatively identical to the non-extremal case, having $\eta = \Lambda^{-1}$ as maximum value for stable field evolution when $\ell = 0$. We underline two specific points in $\eta$,

\begin{align}
\label{qea1}
\eta_\text{I}&=\frac{L^2}{3}-\frac{L^2 m^2 r_h^2 \delta_{+}}{3 n(1+n) r_h^2+3 \ell(\ell+1) \delta_{+}}\ \  \textrm{with}\ \ \delta_{+}=\frac{r_{c}-r_h}{2\kappa_{+}}\textrm{for}\ \ (n\neq0,\ell\neq0)
\end{align}
and
\begin{align}
\eta_\text{II}&=\frac{L^2}{3}+\frac{4 L^2 m^2 r_h^2 \delta_{+}}{3 \left(r_h^2-4 \ell(\ell+1) \delta_{+}\right)}.
\end{align}


Transitions between different regimes of stability can be demonstrated. For $0<\eta<\Lambda^{-1}$ the system is in a stable regime since the imaginary part of the QNM's is negative and constant and the frequencies of oscillation increase rapidly near to $\eta_\text{critical}\sim\Lambda^{-1}$. For $\Lambda^{-1}<\eta<\eta_\text{I}$ the system becomes unstable with a positive purely imaginary QNM which decreases and goes to zero at $\eta_\text{I}$. When $\eta_\text{I}<\eta<\eta_\text{II}$ the system returns to a stable regime with an exponential decay. Beyond $\eta_\text{II}$ the system is still stable but now with an oscillatory exponential decay for late times. In this case, the imaginary part of the QNM's is constant and the frequencies of oscillation tend to a constant.

In the high coupling limit ($\eta\to \infty$) the QNM's are given by
\begin{equation}
\label{omegainfty}
\frac{\omega_{\infty}}{\kappa_{+}}=\sqrt{\ell(\ell+1)-\frac{1}{4}}- i \left(n+\frac{1}{2}\right),
\end{equation}
becoming independent of the mass $m$. In this limit, for $\ell=0$ the modes become purely imaginary $\frac{\omega_{\infty}}{\kappa_{+}}=- i n$, and the fundamental mode $(n=0)$ vanishes.


%


\section{QNM's for non-minimally coupled scalar field evolving in Reissner-Nordstr\"om-de Sitter spacetime}\label{sec4}

In a Reissner-Nordstr\"om-de Sitter spacetime (RNdS), the line-element has exactly the same form of \eqref{metric}, with
\begin{equation}
\l{rn}
f(r) = 1- \frac{2M}{r} + \frac{Q^2}{r^2} - \frac{r^2}{L^2}.
\end{equation}
The functions related to the Klein-Gordon equation as found in \eqref{eqm4} can be written as
\begin{equation}
\l{frn}
\alpha = f^2, \qquad {\Huge  \beta = f \left(1+ \frac{2Q^2 }{r^4}\frac{\eta}{k} \right)} , \qquad \gamma = \frac{f}{k}, \qquad F = \ln \left(kf\right),
\end{equation}
where
\begin{equation}
\l{kk}
 k= k(r, Q, L, \eta )= 1 -\eta A = 1 - \frac{3\eta}{L^2} - \frac{\eta Q^2 }{r^4}.
\end{equation}
In order to separate variables and eliminate the radial first derivative from the wave equation, we must choose the non-trivial field transformation introduced in section \ref{sec1}, namely $R \rightarrow \frac{\tilde{R}}{r\sqrt{k}}$. Together with the tortoise coordinate system, this transformation sets the Klein-Gordon equation in the same simple form of (\ref{eqmstar}). As a drawback, it introduces a discontinuity in the field at the point $k(r)|_{r=r_d}=0$, which poses numerical difficulties. We will then only consider cases where $r_d$ is encapsulated by a horizon and is, therefore, of no consequence.

From the four roots of $f(r)$, at least one is negative, if  two or more real roots exist. We restrict ourselves to the study of a geometry with 3 different horizons, namely, Cauchy $r_y$, event $r_h$ and cosmological horizon $r_c$, with $r_y<r_h<r_c$. As in the Schwarzschild case, the evolution of the scalar field takes place in a region $\mathbb{X}$ defined by $\mathbb{X}:r_h<r<r_c$. Taking $a \equiv \frac{Q}{M}>0$, there are two possible conditions with 3 different positive solutions for $f(r)=0$,\\

\noindent $(i)$ $\dfrac{1}{L^2}<\dfrac{p_+(a)}{32M^2}\quad$ and $\quad a<1$; \hspace{3mm} or\\
$(ii)$ $\dfrac{p_-(a)}{32M^2}<\dfrac{1}{L^2}<\dfrac{p_+(a)}{32M^2}\quad$ and $\quad 1<a<\sqrt{9/8}$,\\

\noindent where $p_{\pm}(a)= \left( -27+36a^2-8a^4 \pm (9-8a^2)^{3/2}\right) /a^6$. The condition $(i)$ recovers the Schwarzschild limit for $a\rightarrow 0$, as stated in \cite{Molina}, with 2 different horizons. The condition $(ii)$ appears as a limit situation in \cite{slovaka}. 
%

\subsection{Effective Potential}

Given the field transformation introduced in section \ref{sec1}. as well as the metric functions defined above, the effective potential for the scalar field reads
\begin{eqnarray}
\nonumber
V(r) = f \left( \frac{2rkk''+4kk'-rk'^2}{4rk^2}f + \frac{2k+rk'}{2rk}f'  + \left(1 + \frac{2Q^2\eta}{r^4k}  \right) \frac{\ell (\ell +1)}{r^2}  + \frac{\mu^2+\xi \mathcal{R}}{k} \right) = \hspace{1.0cm} \\
\nonumber
\frac{f}{k} \left( \frac{2\eta Q^2(9\eta r^4 + \eta L^2 Q^2 - 3L^2r^4)}{r^6(-3\eta r^4 -\eta L^2 Q^2 + L^2r^4)} \left(1- \frac{2M}{r} + \frac{Q^2}{r^2} - \frac{r^2}{L^2}  \right)  \right. \hspace{1.0cm} \\
\left.  +  \left( \frac{\eta Q^2 + r^4 - 3\eta r^4/L^2}{r^5}\right) \left( \frac{2M}{r^2} - \frac{2Q^2}{r^3} - \frac{2r}{L^2}  \right)
 + \left( 1-\frac{3\eta}{L^2} + \frac{\eta Q^2}{r^4}\right)\frac{\ell (\ell +1)}{r^2} + m^2 \right). \hspace{1.0cm}
\l{potrn}
\end{eqnarray}
Again, the Ricci term in $V(r)$ plays no special role in the coupling with the scalar field and is rescaled as previously announced, thus, being not directly related to the presence or absence of unstable modes of the scalar field.

In the limit $Q=0$, the coupled-Schwarzschild potential is recovered. We will not concern ourselves with the point of discontinuity,
\begin{equation}
\l{prd}
r= r_d\big{|}_{k(r)=0} = \left(\frac{\eta Q^2}{1-3\eta / L^2}\right)^{1/4},
\end{equation}
since it originates from our choice of field transformation; we restrain ourselves to one of the ranges: $r_d<r_h$ or $r_c<r_d$, such that $r_d$ is either encapsulated by the event horizon or outside the cosmological horizon. 

\begin{figure}[htbp!]
\begin{center}
\epsfig{file =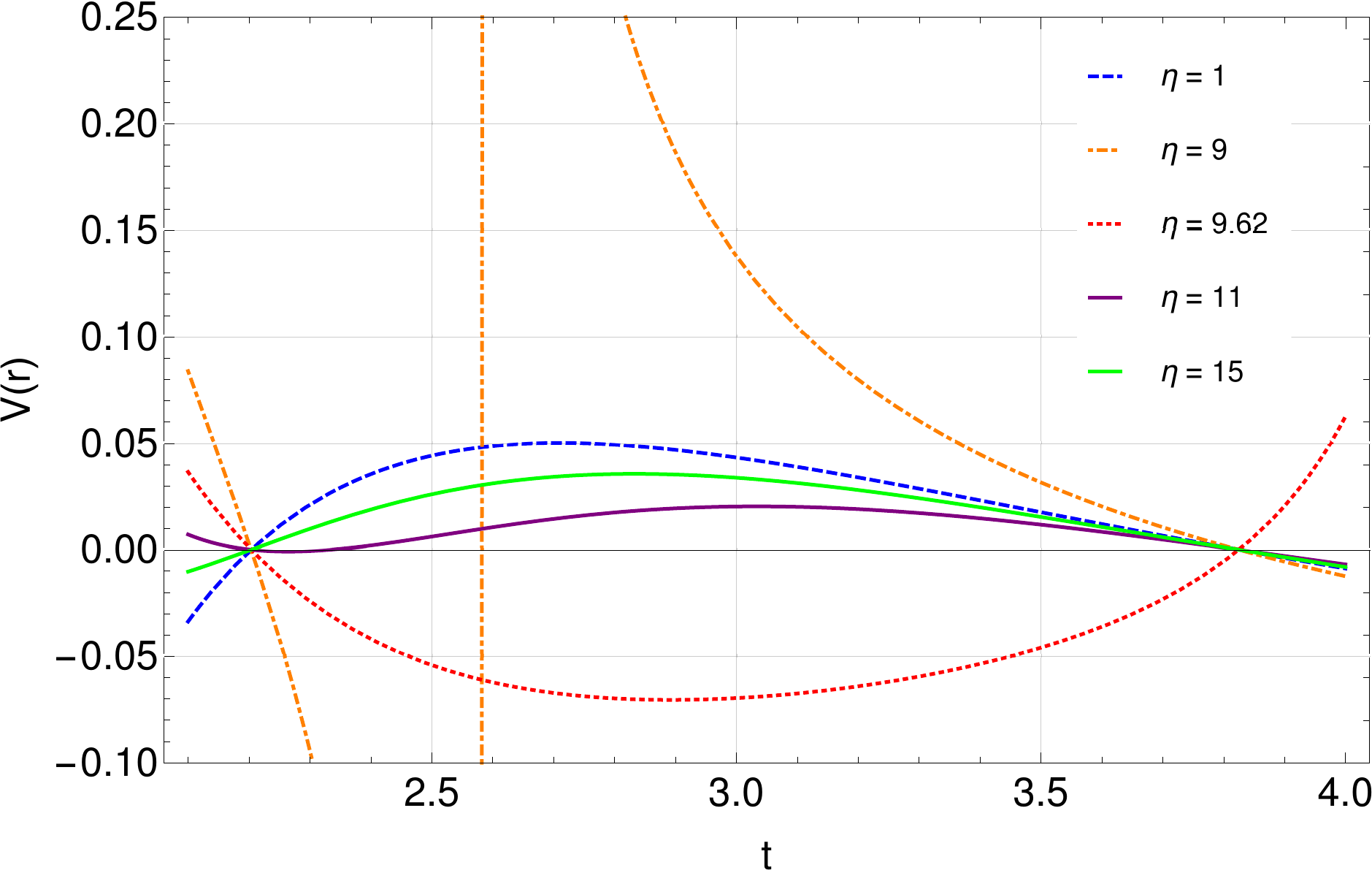, width=6.0cm, height=5.0cm}
\epsfig{file =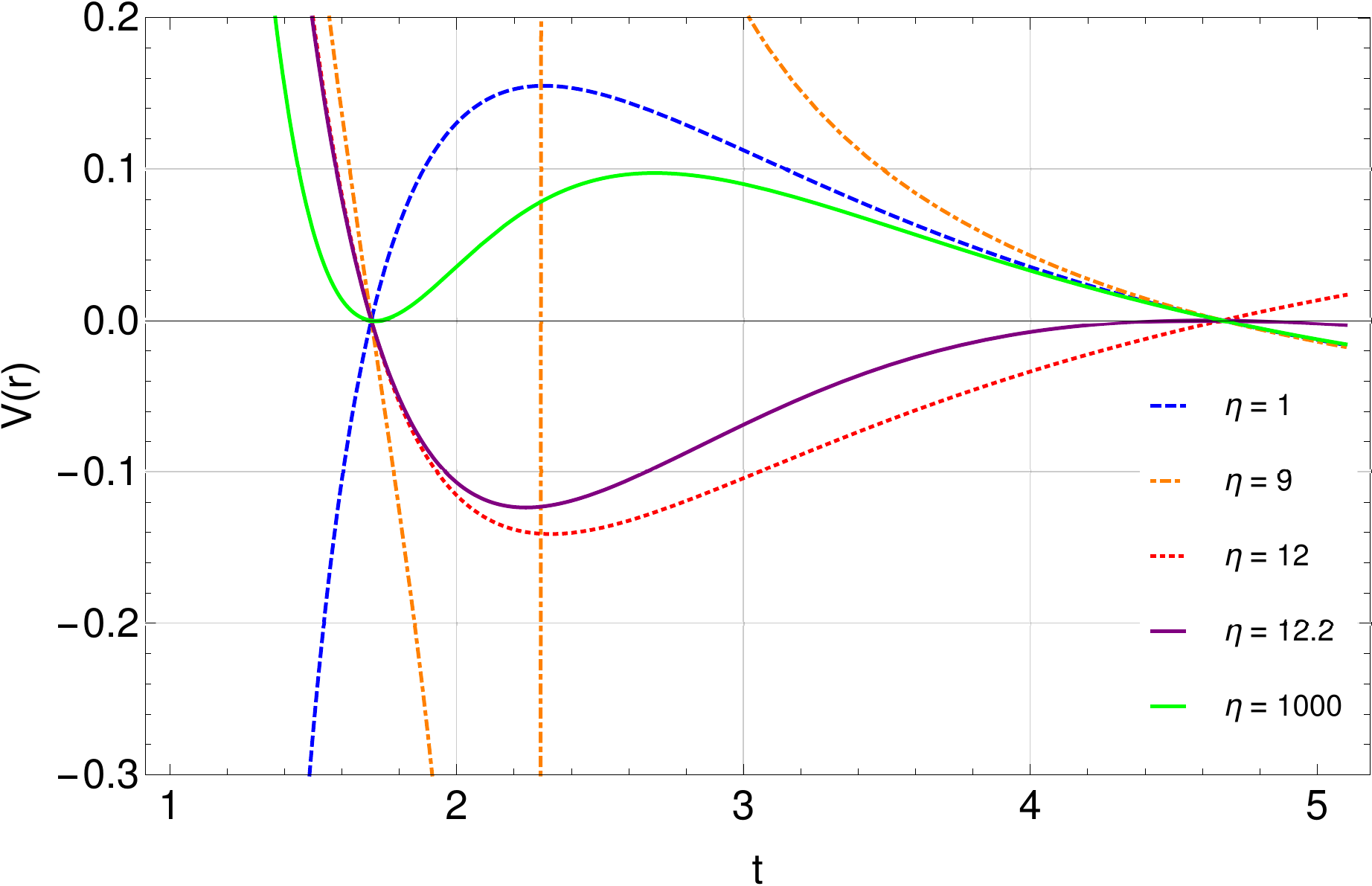, width=6.1cm, height=5.0cm}
\end{center}
\caption{{\small Potential of the Klein-Gordon field in the RNdS geometry with non-minimally coupling. Left panel: Potential for the regions (i) ($\eta = 1$) to (v) ($\eta = 15$) of table~\ref{tab.RNdS1}. The horizons are located at $r_h/2.204 = r_c/ 3.825 = 1$. Right panel: Potential for regions (i) to (iv) of table~\ref{tab.RNdS2}. The two last $\eta$ lie on region (iv). The horizon locations are $r_h/1.703 = r_c/ 4.669 \sim 1$.}}
\label{rnpot}
\end{figure}

Let us consider as an example the massless scalar field, with parameters $M=5Q/3=(L/5.4)^2=\ell /2=1$. In this situation, the potential in $\mathbb{X}$ can be divided, according to its signal, in the five different regions as described in table~\ref{tab.RNdS1}.
A different situation arises, however, when the charge exceeds a critical value $Q_c$. For $M=\ell /2=(L/6)^2=1$, and $m^2=0$, for instance, if $Q > Q_c \sim 0.852$, even for high $\eta$ values, region (v) does not occur. For $Q=0.86$, the potential can be divided in the regions shown in table~\ref{tab.RNdS2}.
Plots for under- and super-critical behaviors are shown in figure \ref{rnpot}.

\begin{table}[ht]
	\caption{Different regions for the potential in RNdS, with $M=5Q/3=(L/5.4)^2=\ell /2=1$.}
	\label{tab.RNdS1}
	\begin{tabular*}{\columnwidth}{*{5}{c@{\extracolsep{\fill}}}}
		\hline
		Case & $\eta$-range & Signal of $V_\mathbb{X} $ & $r_d$ \\
		\hline
		\hline
		(i) & $\eta \lesssim 8.46$ & $V_\mathbb{X}>0$ & $r_d<r_h$ \\
		(ii) & $8.46 \lesssim \eta \lesssim 9.56$ & $V_\mathbb{X}\gtrless0$ & $r_h<r_d<r_c$ \\
		(iii) & $9.56  \lesssim \eta \lesssim 9.88$ & $V_\mathbb{X}<0$ & $r_c<r_d$ \\
		(iv) & $9.88 \lesssim \eta \lesssim 11.52$ & $V_\mathbb{X}\gtrless 0$ & $r_c<r_d$ \\
		(v) & $11.52 \lesssim \eta$  & $V_\mathbb{X}>0$ & $r_c<r_d$ \\
		\hline
	\end{tabular*}
\end{table}


\begin{table}[ht]
\caption{Different regions for the potential in RNdS, for $Q=0.86>Q_c$ (and $M=\ell /2=(L/6)^2=1$; $m^2=0$).}
\label{tab.RNdS2}
\begin{tabular*}{\columnwidth}{*{5}{c@{\extracolsep{\fill}}}}
\hline
Case & $\eta$-range & Signal of $V_\mathbb{X}$ & Place of $r_d$ \\
\hline
\hline
(i) & $\eta \lesssim 5.84$ & $V_\mathbb{X}>0$ & $r_d<r_h$ \\
(ii) & $5.84 \lesssim \eta \lesssim 11.78$ & $V_\mathbb{X}\gtrless 0$ & $r_h<r_d<r_c$ \\
(iii) & $11.78  \lesssim \eta \lesssim 12.23$ & $V_\mathbb{X}<0$ & $r_c<r_d$ \\
(iv) &  $12.23 \lesssim \eta$ & $V_\mathbb{X}\gtrless 0$ & $r_c<r_d$ \\
\hline
\end{tabular*}
\end{table}

The existence of a critical value $Q_c$ is robust against changes in $m$ and $\ell$: for every pair $(\ell ,m )$ when $Q<Q_c$ we can always find a sufficient high $\eta_k$ such that for any $\eta> \eta_k$ we have $V_\mathbb{X}>0$; on the other hand, for $Q>Q_c$ the potential is strictly negative in $\mathbb{X}$.

Considering the different character of the potential for the cases (i) to (v), we can investigate the field evolution by obtaining the system's quasinormal modes and determining whether unstable evolutions are present, or investigate the late time behavior \cite{dsneg} (after the quasinormal ringing). For this reason, we choose to use the characteristic integration over null coordinates to obtain the field profiles together with prony method for the quasi-frequencies. For strictly positive gaussian-like potentials, we compare the frequencies to those obtained with WKB method, with good agreement between the results.

\subsection{Evolution of scalar field: instabilities and QNM's}

\begin{figure}[!ht]
	\begin{center}
		\epsfig{file =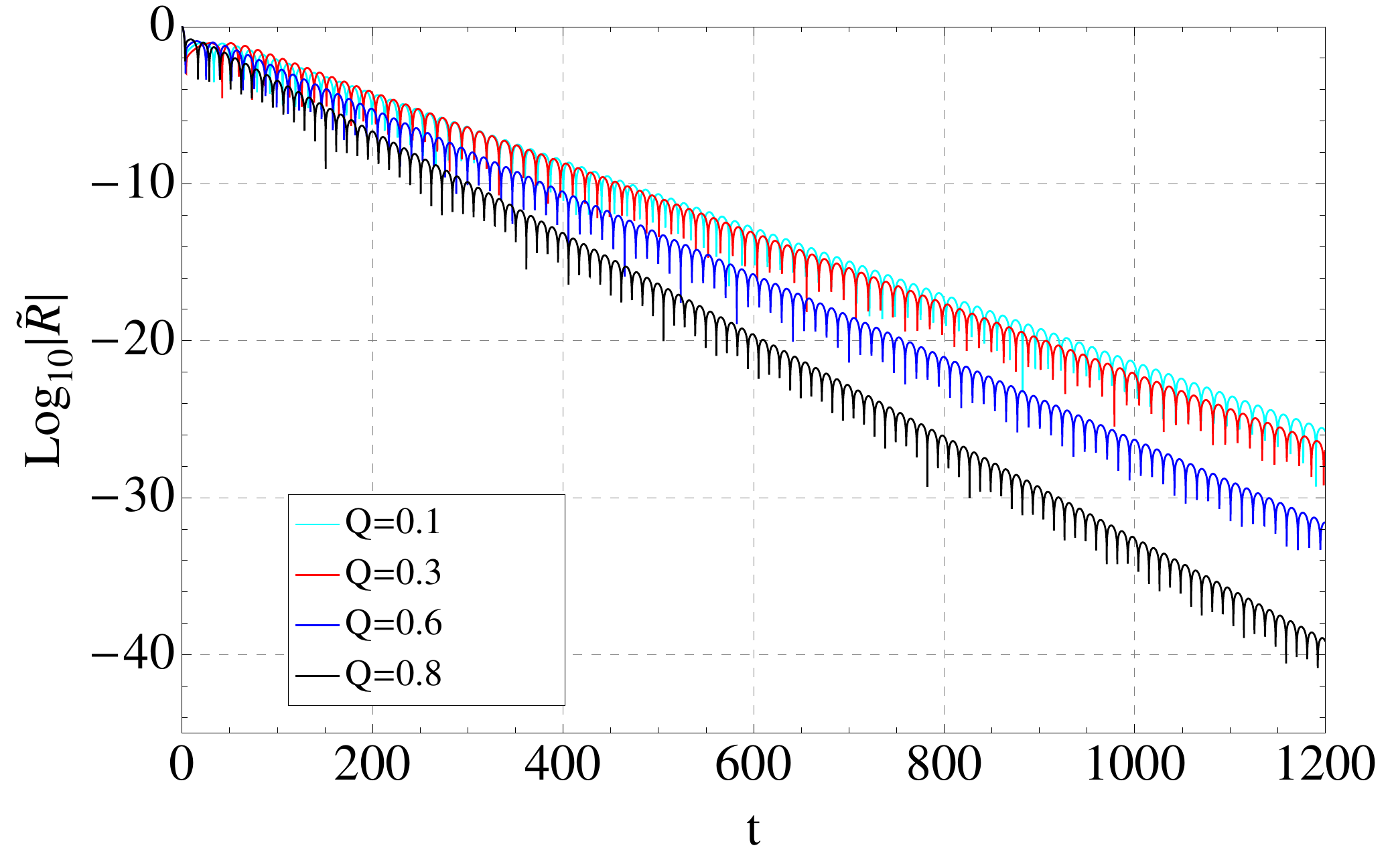, width=6.0cm, height=5.0cm}
		\epsfig{file =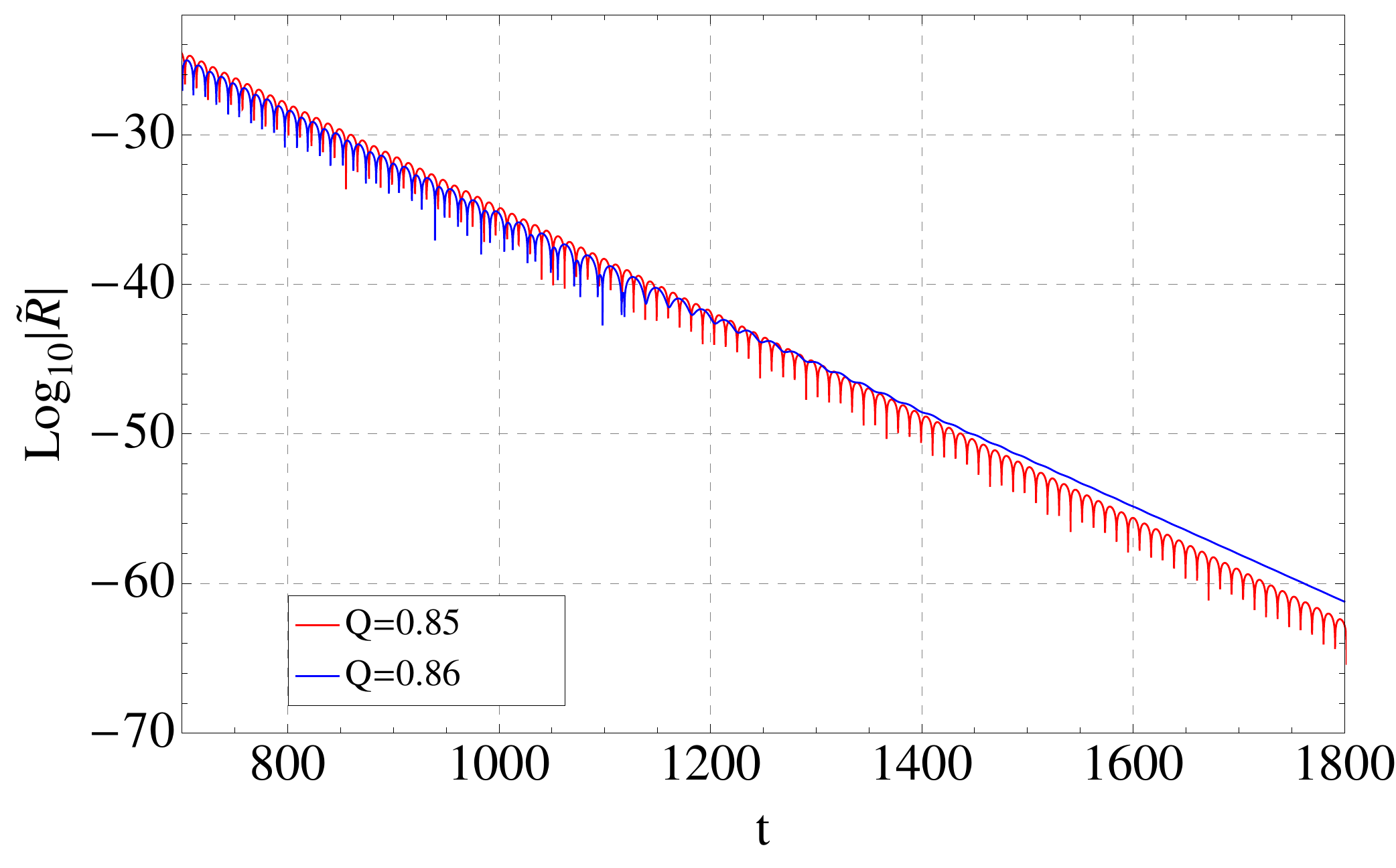, width=6.0cm, height=5.0cm}
		\epsfig{file =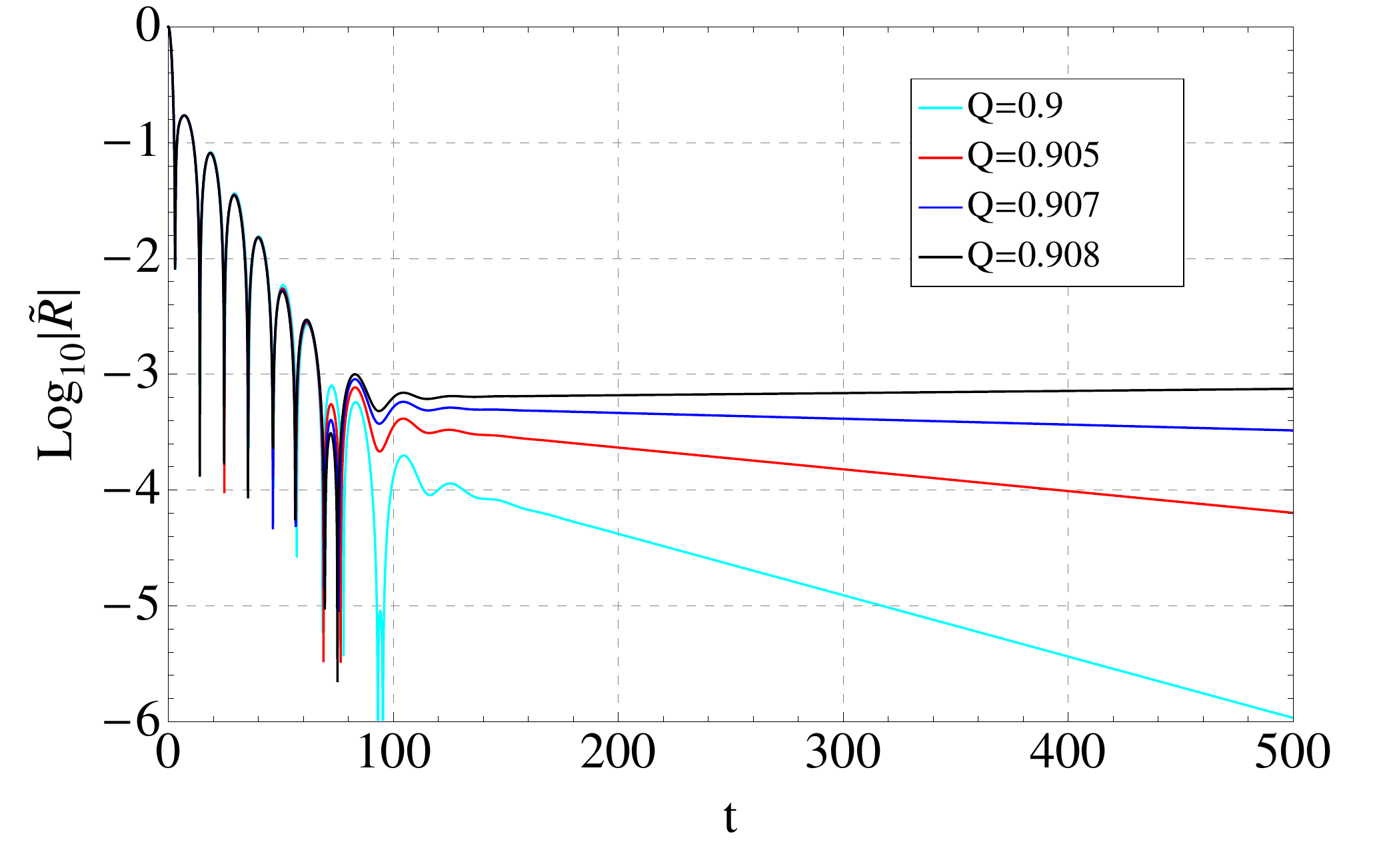, width=6.0cm, height=5.0cm}
		\epsfig{file =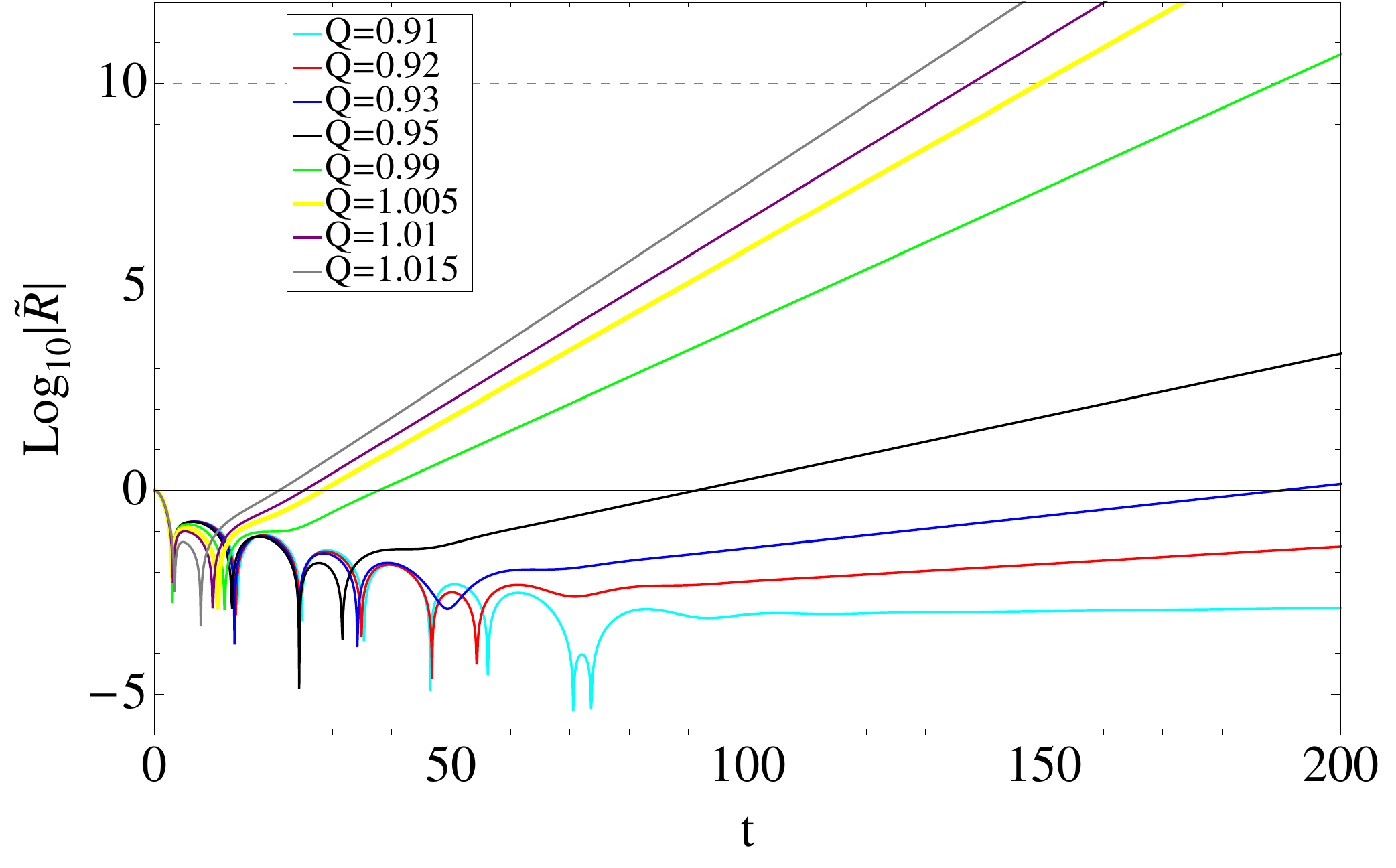, width=6.0cm, height=5.0cm}
	\end{center}
	\caption{{ \small Scalar field propagation in a RNdS black hole with non-minimal coupling with parameters $M=L^2/36=\ell /2 = \eta /50 = 1$, $m^2=0$, resulting in $V_\mathbb{X}>0$ when $Q<Q_c\sim 0.85$. The instabilities in the field arise as a result of the mixed potentials (negative and positive) for large enough $Q$'s and the transition from stable to unstable evolution occurs at some particular charge of the black hole (here $Q \sim 0.908$, in region (iv)). The extremal black hole have $Q=\sqrt{9/8}\sim 1.0156$. }}
	\label{rnds4f}
\end{figure}

The typical evolution of a scalar field coming from the Klein-Gordon equation can be seen in the upper-left panel of figure \ref{rnds4f} for $r=2r_h$ and different values of $Q$. For the chosen parameters, $V_\mathbb{X}>0$ and, as anticipated, there are no instabilities: all profiles decay exponentially in time. Higher values of the black hole charge, however, lead to $V_\mathbb{X}$ partly positive/partly negative, allowing for unstable modes to arise.

%
With the acquired quasinormal signal and the prony method \cite{rosa} we obtain the fundamental quasinormal frequencies up to the critical value of charge $Q_c \sim 0.85$, as listed in the table~\ref{qnmrn_1}.

The obtained frequencies, listed in table \ref{qnmrn_1} are significantly different from the case with no couplings for increasing charge: the $\eta$-coupling is more effective the higher the charge of the black hole, diminishing the rate of increase of $Re (\omega )$ and increasing this rate for the imaginary part.

\begin{table}[ht]
\caption{Fundamental quasinormal modes for non-minimally coupled scalar field evolving in RNdS black holes with different values of $Q$. The spacetime parameters read $M=L/6=\ell /2=\eta /50=1$ and $m^2=0$.The superscript values indicate the deviation of the QNM's from the RNdS case, $\frac{\omega_{RN}-\omega_{\eta}}{\omega_{RN}}$.}
\label{qnmrn_1}
\begin{tabular*}{\columnwidth}{*{7}{c@{\extracolsep{\fill}}}}
\hline
%
%
$Q$ & Re$(\omega )$ &  Im $(\omega )$ && $Q$ & Re$(\omega )$ &  Im $(\omega )$ \\
\hline
\hline
0.05 & $0.2338^{-0.0428\%}$ & $-0.04905^{0.0204\%}$ & & 0.50 & $0.2581^{-6.55\%}$ & $-0.05667^{1.64\%}$ \\
0.10 & $0.2346^{-0.213\%}$ & $-0.04927^{0.0406\%}$ & & 0.55 & $0.2624^{-8.35\%}$ & $-0.05849^{2.36\%}$  \\
0.15 & $0.2360^{-0.466\%}$ & $-0.04963^{0.0605\%}$ & & 0.60 & $0.2667^{-10.5\%}$ & $-0.06064^{3.36\%}$  \\
0.20 & $0.2379^{-0.841\%}$ & $-0.05015^{0.140\%}$ & & 0.65 & $0.2710^{-13.1\%}$ & $-0.06318^{4.75\%}$ \\
0.25 & $0.2402^{-1.37\%}$ & $-0.05081^{0.216\%}$ & & 0.70 & $0.2751^{-16.3\%}$ & $-0.06627^{6.68\%}$ \\
0.30 & $0.2431^{-2.02\%}$ & $-0.05163^{0.349\%}$ & & 0.75 & $0.2792^{-20.0\%}$ & $-0.07003^{9.25\%}$ \\
0.35 & $0.2463^{-2.84\%}$ & $-0.05261^{0.532\%}$ & & 0.80 & $0.2833^{-24.3\%}$ & $-0.07448^{12.7\%}$ \\
0.40 & $0.2500^{-3.84\%}$ & $-0.05377^{0.800\%}$ & & 0.85 & $0.2824^{-29.1\%}$ & $-0.07902^{16.6\%}$ \\
\hline
\end{tabular*}
\end{table}

In figure \ref{rnds4f} we find two field profiles nearby $Q \sim Q_c$ (upper-right panel) and the instabilities found for high values of $Q$ (lower panels). We can see in the same figure (right-bottom panel) the instability of the near extremal black hole to the scalar field for the overcharged black hole ($Q>M$). This is an expected result, given the shape of the potential (very similar to the nearly overcharged black hole, $Q\sim 0.99M$) but is not always the case for every $\eta$: in certain ranges the potential is strictly positive, generating only stable field profiles (e. g. $M=L/6= \ell / 2 = \eta = 1$).

The existence of negative regions in the potential does not ensure the presence of instabilities; otherwise, the negativity on $V_\mathbb{X}$ is related to the presence of an exponential decay in the long-time profile domain. Before $Q_c$, the field oscillates for very long times (right panel in figure \ref{rnds4f}, $Q=0.85$), and beyond this critical charge an exponential decay is shaped as seen in many dS-like geometries \cite{Cardoso, MolinaII} (upper-right panel in figure \ref{rnds4f}).
The exponential decay takes place from $Q=Q_c$ to another high value of $Q$, namely $Q \sim 0.907$ for the assigned parameters (lower panels). For $Q \gtrsim 0.907$, the field growth is unlimited (figure \ref{rnds4f} bottom-left pannel). In this case we may not assume the geometry preserves its original shape: it may evolve to a distinct form. In the right panel on the bottom we see the unstable field evolutions for $M>Q$ to a near-extreme (overcharged) black hole with $\eta$: in every case, the field grows indefinitely showing an unstable behavior.

\begin{figure}[!ht]
\begin{center}
\epsfig{file =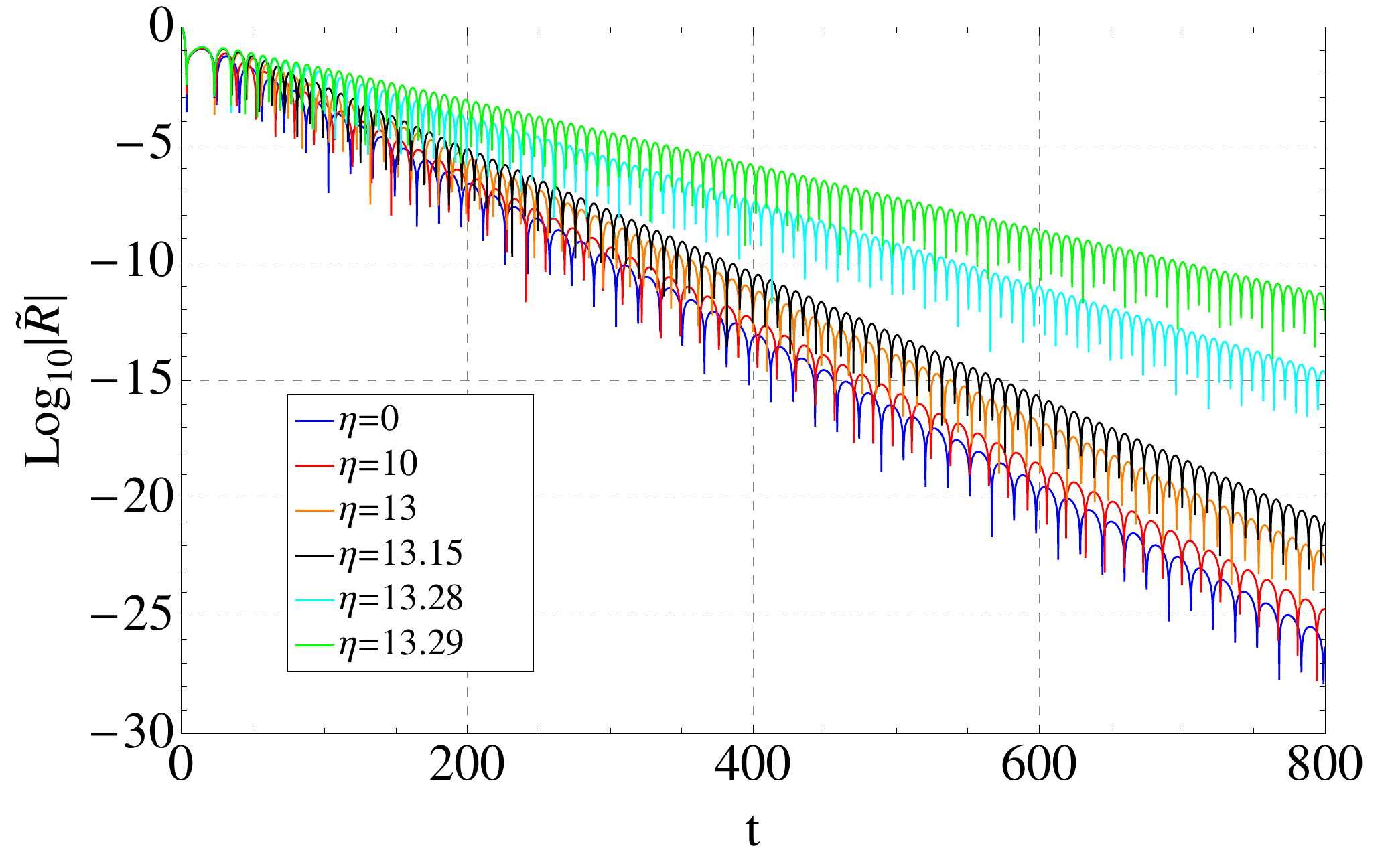, width=6.0cm, height=5.0cm}
\epsfig{file =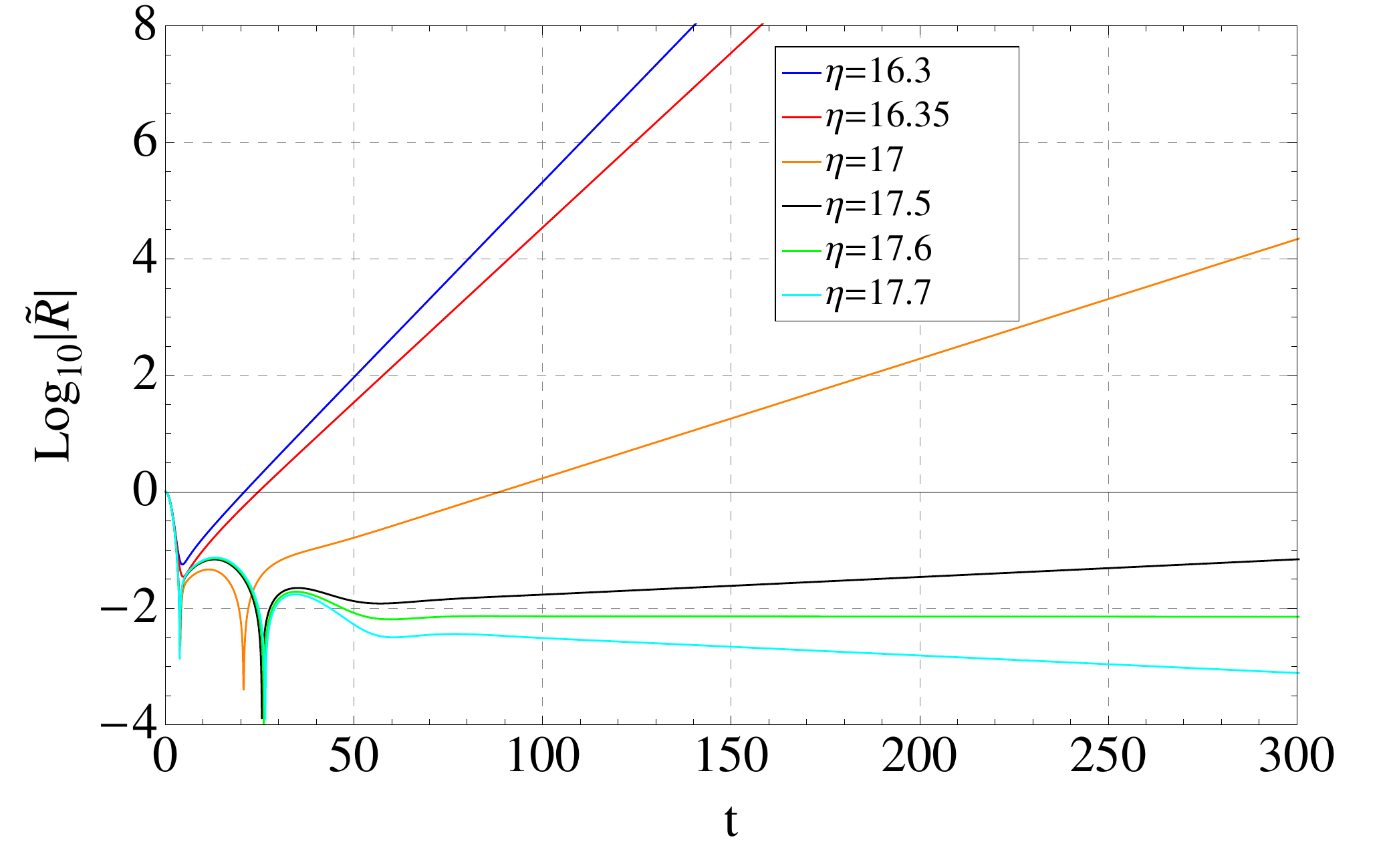, width=6.0cm, height=5.0cm}
\epsfig{file =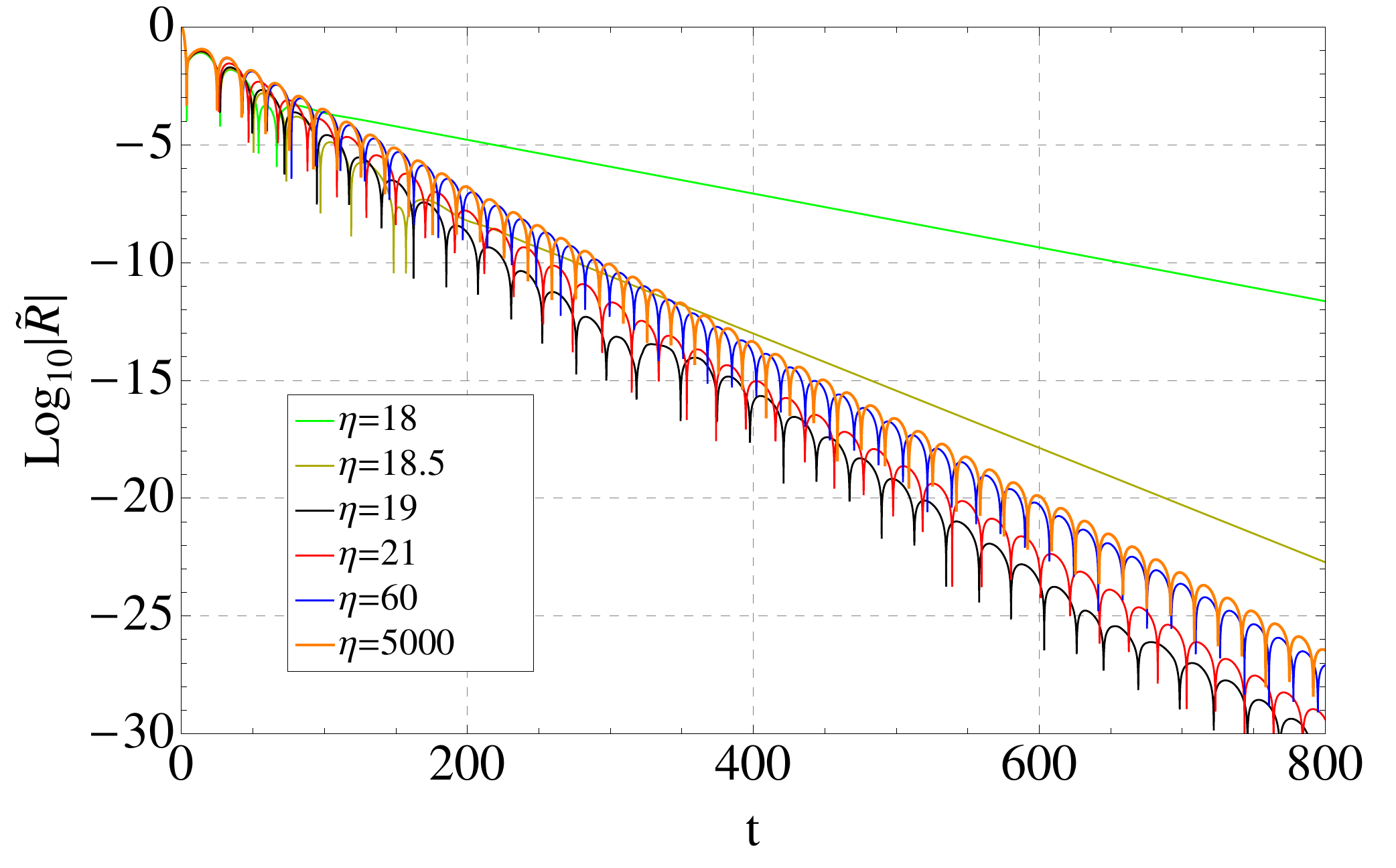, width=6.0cm, height=5.0cm}
\epsfig{file =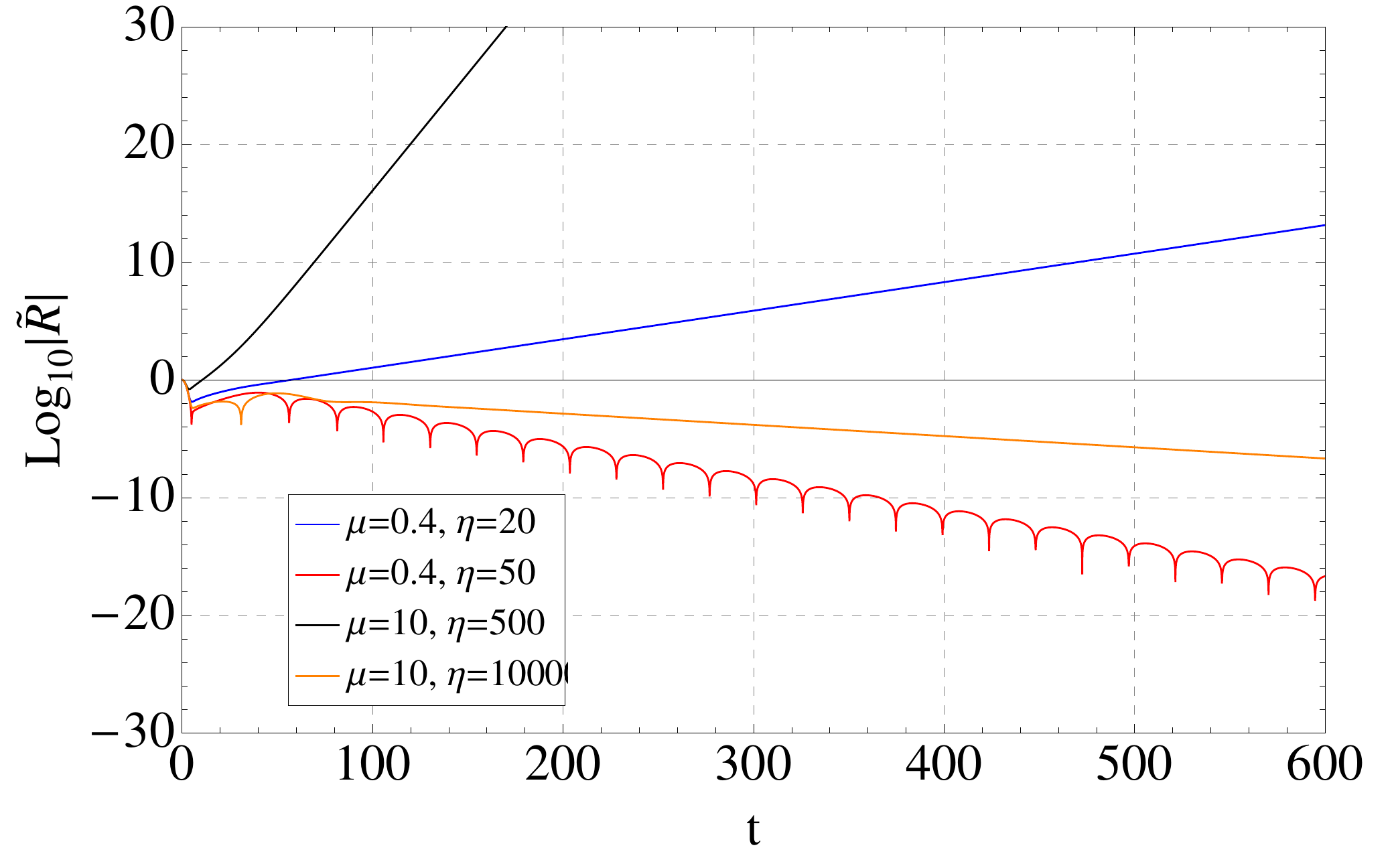, width=6.0cm, height=5.0cm}
\end{center}
\caption{{\small Field profile with parameters $M=\ell =2Q= L^2/49=1$ and $m^2=0$. From upper-left to lower-right, we see the emergence of unstable fields in an intermediate region ((iii) and (iv)) of the $\eta$ parameter. The critical charge related to the formation of region (v) for the chosen parameters is $Q_c \sim 0.777$. In the lower-right panel we see the massive field profile with coupling.}}
\label{rnds5f}
\end{figure}

The presence of a transitional behavior seems to occur also for the variation of $\eta$: by taking fixed $M,Q$ and $L$ we investigate the presence of quasinormal modes and instabilities in regions (i) to (v). In figure \ref{rnds5f} we see different profiles for a large range of $\eta$.

In the upper-left panel of figure \ref{rnds5f} we see the field propagation, with $\eta$-parameter in the region (i), until $\eta^{(1)} \sim 13.29$: the signal damped-oscillates as a never-ending evolution. There are no unstable profiles in this region. From $\eta \sim 13.29$ to $\eta \sim 16.27$ the potential is not continuous (region (ii)). Region (iii), however, is the most critical for the scalar field (since $V_\mathbb{X}<0$ there) and presents the most unstable region. On the upper-right panel we see two profiles with $\eta \sim 16.3$ and $\eta \sim 16.35$, both unstable. The rapidly exponential growth comes from the fact that $V_\mathbb{X}<0$ for almost every $r \in \mathbb{X}$. The fourth region begins at $\eta^{(2)} \sim 16.3985$ going to a maximum value $\eta^{(3)} \sim 21.17$: we can see different field profiles in the upper-right and lower-left panels stable/unstable, depending on the parameter. The transition from stable to unstable profiles takes place nearby $\eta \sim 17.6$, still in region four: both kinds of signal occur, depending on {\it{how negative}} the potential is. After $\eta \sim 21.17$ (lower left-panel) we see the last two quasinormal modes (region (v)), as never-ending damping oscillating signals. In the high $\eta$ regime, the field profile approaches the absent coupling case and oscillates following closely the Reissner-Nordstr\"om-de Sitter record.

\begin{table}[ht]
\caption{Fundamental quasinormal modes for non-minimally coupled scalar field evolving in RNdS black holes with varying $\eta$. The spacetime parameters read $M=L/7=\ell =2Q=1$ and $m^2=0$.}
\label{qnmrn_2}
\begin{tabular*}{\columnwidth}{*{7}{c@{\extracolsep{\fill}}}}
\hline
$\eta$ & Re$(\omega )$ &  Im $(\omega )$ && $\eta$ & Re$(\omega )$ &  Im $(\omega )$ \\
\hline
\hline
$0$ & $0.2034$ & $-0.07374$ & & $13.29$ & $0.4258$ & $-0.03163$ \\
$3$ & $0.2067$ & $-0.07308$ & & $13.2935$ & $0.4389$ & $-0.01148$  \\
$7$ & $0.2159$ & $-0.07213$ & & $19$ & $0.1407$ & $-0.08782$  \\
$10$ & $0.2328$ & $-0.07074$ & & $21$ & $0.1524$ & $-0.08197$ \\
$11$ & $0.2449$ & $-0.07008$ & & $60$ & $0.1835$ & $-0.07739$ \\
$13$ & $0.3287$ & $-0.06457$ & & $5000$ & $0.1884$ & $-0.07569$ \\
$13.15$ & $0.3553$ & $-0.06061$ & & $10^9$ & $0.1884$ & $-0.07568$ \\
$13.28$ & $0.4115$ & $-0.04171$ & &  \multicolumn{3}{c}{} \\
\hline
\end{tabular*}
\end{table}

In table \ref{qnmrn_2} it is make clear the influence of varying-$\eta$ in the scalar field propagation: for the first region of the potential, the higher the $\eta$, the higher the quality factor of the black hole\footnote{Re$(\omega )$ increases and Im$(\omega )$ decreases.}. We must be attentive still, of the high variation when getting closer to the frontier of (i) in $\eta$: from $\eta = 13.29$ to $\eta = 13.2935$ we have a $\Delta \eta \rightarrow 0.026\%$ variation whereas $\Delta\omega_{R} \rightarrow 3.1\%$ and $\Delta\omega_{I} \rightarrow 64\%$. This type of change characterizes a variation similar to that occurred in the near extremal regime (when the accretion of small amounts of charge in the black hole induces huge variations in the spectra of the oscillation).

Another interesting picture in the quasinormal spectrum with NMDC is the existence of an asymptotic value of $\omega$ for high $\eta$: in the table we can see, to the 4 figure, the QNM is the same for $\eta = 5000$ and $\eta = 10^9$, both cases in region (v). The last feature we emphasize, is the highest values of Im$(\omega )$ and the smallest for Re$(\omega )$ both in region (iv). This is an expected feature in relation to the imaginary part, as long as regions (ii), (iii) and (iv) are the unstable ones.

Considering the field evolution of $\ell>0$ profiles for different $\eta$, the general behavior in the potential is the raising of its maxima/minima points, which does not relate to the formation of stable asymptotic regions (region (v)). Otherwise, this fact is related to the range at which we have unstable modes: the higher the multipole number, the less stable the scalar field tends to perform. Still, the range of stability in $\eta$ diminishes for increasing $\ell$: in table \ref{cr1} we list critical values for the coupling for which the field evolves stably; whenever $\eta < 13.2935$ or $\eta > \eta_c$, the field is stable.

\begin{table}[ht]
\caption{Critical values of $\eta$ (the field evolves stably after $\eta > \eta_c$.}
\label{cr1}
\begin{tabular*}{\columnwidth}{*{9}{c@{\extracolsep{\fill}}}}
\hline
\multicolumn{9}{c}{$M=2Q=L/7=1$ and $m^2=0$} \\
\hline
\hline
$\ell$ & 0 & 1 & 2 & 3 & 4 & 5 & 10 & 20 \\
\hline
$\eta_c$ & $16.25^{\pm 0.05}$ & $17.55^{\pm 0.05}$ & $18.25^{\pm 0.05}$ & $18.75^{\pm 0.05}$ & $19.05^{\pm 0.05}$ & $19.35^{\pm 0.05}$& $20.05^{\pm 0.05}$ & $20.35^{\pm 0.05}$ \\
\hline
\end{tabular*}
\end{table}

From the same table, we realize that $\eta_c$ increases for increasing $\ell$. Possibly the values of $\eta_c$ approach a finite asymptote when we take $\ell \rightarrow \infty$, given the growing of the $\ell$ x $\eta_c$ curve, what is not possible to be investigated numerically\footnote{The higher the $\ell$, the higher the time (in the field) to which we must integrate in order to obtain the exponential growth/decay of the field. This represents a geometrical growth in time of computation versus an arithmetic growth in $\ell$}.

As in the de Sitter geometries with black holes \cite{Molina} the scalar field multipole $\ell=0$ is a special case. Although not conclusive, for late times, the field tends to increase very slowly to a constant value (for very late times). For instance, taking $M=Q/2=\eta/500=L/7=1$, the evolution seems to evolve very slowly to an asymptotic value ($\tilde{R}\sim -0.03$), for late times.

In general, for all multipole number, we demonstrate the presence of a gap of instability in $\eta$-range for the scalar field: when $\eta < 13.2935$ or for $\eta > \eta_c$ the field evolves stably, being unstable if $\eta < \eta_c$, in regions (iii) and (iv) (as stated before, it is not possible to obtain numerical integration in region (ii)).

In the regime of high cosmological radius, we can see the formation of region (iii) and (iv) in the potential when $\eta > \frac{L^2}{3}$, but no region (v) as a general feature. Even for small values of charge, we have no region (v), but the gap for existence of region (iii) is very small in $\eta$. As an example, let us assume a geometry with $M=100Q=\ell =1$, $m =0$ and $L=6\times 10^{5}$.  The region (iii) for negative potential almost vanishes: it endures a range of $\Delta \eta \sim 10^{-10}$, after the critical $\eta \rightarrow \Lambda^{-1}$ (for $Q=1/2$ and the same parameters, $\Delta \eta \sim 10^{-6}$). The region after that, region (iv) appears for every $\eta$. Region (v) will only emerge in cases with very small black hole charges ($Q<Q_c \sim 10^{-5}$), for example nearby $\eta \sim 4\Lambda^{-1}$. The general behavior remains, however: at some point for each geometry, we will have the critical value of $\eta$ from which the field evolves stably.

The situation changes drastically, though, if we add a small scalar field mass to the last scenario. Taking $m \sim 0.1$, as an example,  we have a range $\Delta \eta \sim 10^9$ of unstable fields after the point $\eta^{(2)} \sim 1.2 \times 10^9$. In figure \ref{rnds5f} in lower-right panel we see the coupling of $\eta$ and the scalar field: in general, the higher the scalar field mass, the higher the value of $\eta$ for the formation of a stable region of oscillations in the potential.

In the specific case when $\eta = \frac{L^2}{3}$, the field equation may be evolved with a simpler potential than Eq. \ref{potrn}, 
\begin{align}
\label{ext1}
V(r)= f \left[\frac{f'}{r} +\frac{\ell ( \ell + 1)}{r^2} -\frac{2f}{r^2} \right]
\end{align}
The proper field transformation for this coupling is given by $R\rightarrow \tilde{R} \frac{1}{r\sqrt{k}} \rightarrow \tilde{R} r$ and when we apply it to the scalar equation with the tortoise coordinate, it brings (\ref{ext1}) as a result. 

The potential implies unstable scalar evolution, what can be seen by analyzing the extra term, $2f/r^2$: it is always positive in the region of field propagation. In such case, the wave equation is the same as that for the scalar field in Reissner-Nordstr\"om-de Sitter geometry, with a negative term inside the brackets. Even though this term varies with $r$, the fact that it is always negative is sufficient to assure the unstable evolution of the scalar field whenever $\ell = 0$:  the field propagating in RNdS geometry with negative square mass (even of very small masses) is unstable (the same result being true in our case). This result is very similar to that of the Schwarzschild section: the instability for $\eta > L^2/3$, comes as a result of the effective mass being negative in that limit.

The numerical data obtained by evolving the scalar filed in the potential (\ref{ext1}) turns out unstable in all tested parameters, for $M=1$, $\ell = 0$, $L=7$ and $L=50$, and $Q=0.01, 0.1, 1.001$ (as expected).
\\

{\it Quasi-extremal regime.}
\\

The quasi-extremal regime in a RNdS black hole has two possible horizon coalescence, $r_y=r_h$ (high $Q$) or $r_c=r_h$ (high $\Lambda$). In the first case, given the high values of charges, region (v) never exists. In this case, all the tested profiles of region (iii) and (iv) for $\ell$ come out stable whenever $\ell >0$. On the other hand, taking for example, $M=L/6=1$ and $\delta \equiv \frac{Q_{ext}-Q}{Q_{ext}} \sim 10^{-9}$, for asymptotic $\eta$, the scalar field turns out unstable. The potential forms region (iii) for $11.7<\eta < 12.3$, but, as long as all field profiles evolve unstably in region (v), $\eta > 11.7$ represents an unstable range of parameter. This was tested for multiple $\eta$ and $\ell =1$, but can be also take as granted for other $\ell>1$ as long as the deep of the potential grows in those cases. Again we have most probably a stable evolution for $\ell =0$, qualitatively similar to the one discussed in the previous subsection for the non-extremal case. In that way we can still assure the presence of region (i) in the potential when $\eta < 2.22$, and the field evolves stably as a quasi-oscillation or an exponential decay after the initial burst.

When the cosmological constant is high, we have a more interesting frame. If we take, for instance, $M=5Q/3=1$ and $L=4.8587$ ($\Lambda = 0.999998\Lambda_{ext}$), regions (ii) and (iv) happen for very small intervals in $\eta$ of order of $10^{-4}$. In such case, the evidence of a gap of instability is very pronounced. For $\eta < 7.49$ or $\eta > 8.29$, the field evolves stably for every $\ell$\footnote{Regions (ii) and (iv) form around $(7.4895, 7.4921)$ and $(8.285, 8.289)$, respectively.}.

Repeating the same procedure declared in the previous section we calculated the QNM's for massive NMDC scalar field in quasi-extremal limit $(r_y \sim r_h)$  in terms of $(\ell,m,\eta)$ and the characteristic parameters of the black holes $(r_h,r_{y},\kappa_{+})$ as follows
\begin{equation}
\label{qnmextremalrn}
\frac{\omega}{\kappa_{+}}=\sqrt{\left(\frac{\ell(\ell+1)}{r_h^{2}}\left[\frac{L^2(r^2_{h}+\eta)-6r^2_{h} \eta}{L^2 (r^2_h -\eta)}\right]+\frac{m^2 r^2_h}{r^2_h -\eta}\right)\frac{r_{y}-r_h}{2\kappa_{+}}-\frac{1}{4}}- i \left(n+\frac{1}{2}\right).
\end{equation}
Now the range of critical values of $\eta$ is more intricate. Differently from Schwarzschild-de Sitter case, here we can have unstable modes even if $m=0$ when $\eta>\frac{L^2r_h^2}{6 r_h^2-L^2}$. By taking $\ell=0$ (the most interesting case), we have
\begin{equation}
\label{qnmextremalrnl0}
\frac{\omega}{\kappa_{+}}=\sqrt{\left(\frac{m^2 r^2_h}{r^2_h -\eta}\right)\frac{r_{y}-r_h}{2\kappa_{+}}-\frac{1}{4}}- i \left(n+\frac{1}{2}\right).
\end{equation}
Inspecting the Eq.(\ref{qnmextremalrnl0}) one can see that the critical value is $\eta=r_h^2$ and now the two specific points are
\begin{equation}
\label{etanonrealell0}
\eta_I =r_h^2\left(1 - 4 \mu^2 \delta_+\right),
\end{equation}
and
\begin{eqnarray}
\label{etaunstable}
\eta_{II}=r_h^2 \left(1+\frac{\mu^2\delta_+}{n(n+1)}\right).
\end{eqnarray}
The stability condition of these quasinormal modes are the same of presented for Schwarzschild-de Sitter. In the high coupling limit ($\eta\to \infty$) the QNM's are given by
\begin{equation}
\label{omegainftyrnds}
\frac{\omega_{\infty}}{\kappa_{+}}=\sqrt{-\left(\frac{\ell(\ell+1)}{r_h^{2}}\left[1-\frac{6 r^2_h}{L^2}\right]\right)\frac{r_{y}-r_h}{2\kappa_{+}}-\frac{1}{4}}- i \left(n+\frac{1}{2}\right),
\end{equation}
becoming independent of the mass $m$. In this limit, differently of Schwarzschild-de Sitter, unstable modes will be present if $\Lambda^{-1}>2 r^2_h$.  



\section{Final remarks}\label{sec6}
In the present work we discussed the effect of a non-minimally derivative coupling on the dynamics of a scalar field propagating in asymptotically dS spacetimes. Three different cases were studied: the de Sitter, Schwarzschild-de Sitter, and Reissner-Nordstr\"om-de Sitter metrics.

Considering the evolution of a scalar probe field in a four-dimensional dS spacetime, we computed the quasinormal spectrum when the NMDC term $\eta$ is present. We found growing quasinormal modes in the positive branch of frequencies leading to regions of instability. In the context of dS/CFT correspondence, we generalize the result for the two-point Hadammad function, showing that its poles match with the regions of instability in the quasinormal spectrum.

In the case of Schwarzschild-de Sitter geometry, the presence of an $\eta$-term in the field equation also introduces instabilities in the quasinormal spectra for a given range of $\eta$. The effect of the coupling is to modify the square mass of the scalar field turning it negative in certain ranges of values, presenting expected instabilities for the field evolution ($\ell=0$). 

The cases with low values of $\ell$ are the most unstable, numerically. In particular for $\eta > \Lambda^{-1}$ the field becomes unstable ($\ell=0$). For different angular momentum, though $\ell >0$, the profile turns out stable after a transitional value $\eta_T$,

\begin{equation}
\label{fr1}
\eta_T \sim \frac{L^2}{3}\left( 1 + \frac{m^2L^2}{\ell(\ell+3)} \right)
\end{equation}
for small $\Lambda$. The expression above is very similar to \eqref{qea1} for $n=0$ and $L \sim r_h$, in the quasi-extremal regime.

Surprisingly, the massless scalar field equation is not affected by the coupling. The spectra of frequencies is stable, as expressed by the usual Schwarzschild-de Sitter quasinormal modes. This was shown to be the case also in the de Sitter spacetime.

The same is not true for the Reissner-Nordstr\"om black hole in a dS geometry, where even the massless scalar field is affected for the non-minimally coupling constant. The potential is significantly more complicated, compared to the chargeless case, possessing five qualitative different regions according to its sign. In regions (iii) and (iv) we have two critical constants, $\eta^{(2)}$ and $\eta^{(3)}$, determined by the spacetime parameters, such that for $\eta^{(2)}< \eta < \eta^{(3)}$ unstable modes are present. The range of $\eta$ for which unstable modes are present grows as we increase the charge of the black hole.

A range of instability for $\eta$ occurs for every $\ell$ (differently from \cite{Chen}). The frequencies are sensitive to the variation of the $\eta$-parameter, being the quasinormal spectrum particularly affected by its presence. 

For every $\eta$ it is always possible to find a range of charges of the black hole for which unstable modes are present, suggesting $\eta$ might be an appropriate order parameter for studying critical phenomena in these systems.

In the quasi-extremal limit for Schwarzschild-de Sitter and Reissner-Nordstr\"om-de Sitter, the quasinormal spectra was obtained exactly, following the approach of \cite{MolinaII,Cardoso}, and the observed behavior is similar to that of the non-quasi-extremal case.

The investigation of the presence of instabilities is a fruitful field of research. In this work, the peculiar evolution of a probe scalar field in a number of geometries revealed critical phenomena which may be related to second order phase transitions present in the corresponding CFT side of theory. Non-minimally coupled models enable a vast amount of dynamical field analysis, with parameter ranges over which the spacetime is unstable being a particularly important feature.

\begin{acknowledgements}
The authors would like thank Jefferson Stafusa Elias Portela for critical comments to the manuscript. This work was supported by CNPq (Conselho Nacional de Desenvolvimento Cient\'{\i}fico e Tecnol\'ogico), FAPESP (Funda\c c\~ao de Amparo \`a Pesquisa do Estado de S\~ao Paulo) and FAPEMIG (Funda\c c\~ao de Amparo \`a Pesquisa do Estado de Minas Gerais), Brazil.
\end{acknowledgements}

\appendix

\section{Einstein's tensor components and Ricci scalar for a four-dimensional spherically symmetric spacetime}\label{app}

The Einstein tensor $G^{\mu\nu}$ has only the diagonal components that can be written in terms of the components of the metric and functions $A(r)$ and $B(r)$. The non-vanishing components of the Einstein tensor for a spherically symmetric static spacetime can be put in a general form
\begin{align}
G^{tt}=\frac{A}{f}&=\frac{1}{f}\left( \frac{(1-f)}{r^2}-\frac{f'}{r}\right) \\
\nonumber\\
G^{rr}=-A f&=-f \left( \frac{(1-f)}{r^2}-\frac{f'}{r}\right) \\
\nonumber\\
G^{\theta\theta}&=\frac{B}{r^2}=\frac{1}{r^2}\left(\frac{f'}{r}+\frac{f''}{2}\right)=\sin^2 \theta G^{\phi\phi},
\end{align}
and the Ricci scalar is given by
\begin{align}
\mathcal{R}=-\left(f''+\frac{4f'}{r}+\frac{2(f-1)}{r^2}\right).
\end{align}

\subsection{RNdS}

In this case, the Einstein tensor in a covariant-form is expressed as
\begin{displaymath}
\label{sds3}
G^{\mu\nu}=G^{\mu\nu}_{\Lambda}+G^{\mu\nu}_{EM} = -\frac{3}{L^2}\left[\begin{array}{cccc}
-\frac{1}{f(r)}& 0 & 0 &0\\
0 & f(r) & 0& 0\\
0 &0&\frac{1}{r^2}& 0\\
0 &0&0&\frac{1}{r^2\sin^2\theta}
\end{array}\right] +\frac{Q^2}{r^4}\left[\begin{array}{cccc}
\frac{1}{f(r)}& 0 & 0 &0\\
0 & -f(r) & 0& 0\\
0 &0&\frac{1}{r^2}& 0\\
0 &0&0&\frac{1}{r^2\sin^2\theta}
\end{array}\right]
\end{displaymath}
and the remaining relations for $A$ and $B$ are $A=\frac{3}{L^2}+ \frac{Q^2}{r^4}$ and $B=-\frac{3}{L^2}+ \frac{Q^2}{r^4}$. In the Schwarzschild case, we can easily obtain the same relations by taking $Q=0$, (then $A=-B$).

In four dimensions the Ricci scalar for the RNdS spacetime has the same value as for the SdS spacetime,
\begin{align}
\mathcal{R}=\frac{12}{L^2}.
\end{align}



\begin{thebibliography}{99}

\bibitem{Berti:2009kk}
 E.~Berti, V.~Cardoso, A.~O.~Starinets,
  Class.\ Quant.\ Grav.\  {\bf 26} (2009) 163001.
  doi:10.1088/0264-9381/26/16/163001.
  arXiv:09052975 [gr-qc]


\bibitem{Kokkotas:1999bd}
   K.~D.~Kokkotas, B.~G.~Schmidt,
  Living Rev.\ Rel.\  {\bf 2} (1999) 2.
  doi:10.12942/lrr-1999-2.
  arXiv:9909058 [gr-qc].

\bibitem{Nollert:1999ji}
  H.~P.~Nollert,
  Class.\ Quant.\ Grav.\  {\bf 16} (1999) R159.
  doi:10.1088/0264-9381/16/12/201.

\bibitem{Abbott:2016blz}
  B.~P.~Abbott {\it et al.} [LIGO Scientific and Virgo Collaborations],
  Phys.\ Rev.\ Lett.\  {\bf 116}, no. 6, (2016) 061102.
  doi:10.1103/PhysRevLett.116.061102.
  arXiv:1602.03837 [gr-qc].


\bibitem{TheLIGOScientific:2017qsa}
  B.~P.~Abbott {\it et al.} [LIGO Scientific and Virgo Collaborations],
  Phys.\ Rev.\ Lett.\  {\bf 119}, no. 16, (2017) 161101.
  doi:10.1103/PhysRevLett.119.161101.
  arXiv:1710.05832 [gr-qc]

\bibitem{Regge:1957td}
  T.~Regge, J.~A.~Wheeler,
  Phys.\ Rev.\  {\bf 108}, (1957) 1063.
  doi:10.1103/PhysRev.108.1063.

\bibitem{Nunez:2003eq}
  A.~Nunez, A.~O.~Starinets,
  Phys.\ Rev.\ D {\bf 67}, (2003) 124013. doi:10.1103/PhysRevD.67.124013.
  arXiv:0302026 [hep-th]

\bibitem{Horowitz:1999jd}
  G.~T.~Horowitz, V.~E.~Hubeny,
  Phys.\ Rev.\ D {\bf 62}, (2000) 024027.
  doi:10.1103/PhysRevD.62.024027.
  arXiv: 9909056 [hep-th]

\bibitem{Hartnoll:2009sz}
  S.~A.~Hartnoll,
  Class.\ Quant.\ Grav.\  {\bf 26}, (2009) 224002.
  doi:10.1088/0264-9381/26/22/224002.
  arXiv:0903.3246 [hep-th]

\bibitem{Son:2002sd}
  D.~T.~Son, A.~O.~Starinets,
  JHEP {\bf 0209}, (2002) 042.
  doi:10.1088/1126-6708/2002/09/042.
  arXiv:0205051 [hep-th]


\bibitem{Sybesma:2015oha}
  W.~Sybesma, S.~Vandoren,
  JHEP {\bf 1505}, (2015) 021.
  doi:10.1007/JHEP05(2015)021.
  arXiv:1503.07457 [hep-th]

\bibitem{elciobinlimaqiu}E.~Abdalla, B.~Wang, A.~Lima-Santos, W.~G.~Qiu,
  Phys.\ Lett.\ B {\bf 538},(2002) 435.
  [Conf.\ Proc.\ C {\bf 0208124},(2002) 322].
  doi:10.1016/S0370-2693(02)02039-7.
  arXiv:0204030 [hep-th]

  \bibitem{elciokarluciolima}
  E.~Abdalla, K.~H.~C.~Castello-Branco, A.~Lima-Santos,
  Phys.\ Rev.\ D {\bf 66},(2002) 104018.
  doi:10.1103/PhysRevD.66.104018.
  arXiv:0208065 [hep-th]

  \bibitem{chandrasekhar} S. Chandrasekhar, \textit{The Mathematical Theory of 	 Black Holes}, (Clarendon Press, Oxford UK 1985), 646 p.

  \bibitem{Bekenstein:1996pn}
  J.~D.~Bekenstein,
  ``Black hole hair: 25 - years after,''
  In *Moscow 1996, 2nd International A.D. Sakharov Conference on physics*     	216-219.
  arXiv:9605059[gr-qc]

  \bibitem{Gubser:2008px}
  S.~S.~Gubser,
  Phys.\ Rev.\ D {\bf 78}, (2008) 065034.
  doi:10.1103/PhysRevD.78.065034.
  arXiv:0801.2977 [hep-th].

  \bibitem{Rinaldi:2012vy}
  M.~Rinaldi,
  Phys.\ Rev.\ D {\bf 86}, (2012) 084048.
  doi:10.1103/PhysRevD.86.084048.
  arXiv:1208.0103 [gr-qc]

  \bibitem{Minamitsuji:2013ura}
  M.~Minamitsuji,
  Phys.\ Rev.\ D {\bf 89}, (2014) 064017.
  doi:10.1103/PhysRevD.89.064017.
  arXiv:1312.3759 [gr-qc]

  \bibitem{Volkov:1998cc}
  M.~S.~Volkov, D.~V.~Gal'tsov,
  Phys.\ Rept.\  {\bf 319}, (1999) 1.
  doi:10.1016/S0370-1573(99)00010-1.
  arXiv:9810070 [hep-th]

\bibitem{Hartnoll:2008kx}
  S.~A.~Hartnoll, C.~P.~Herzog, G.~T.~Horowitz,
  JHEP {\bf 0812}, (2008) 015.
  doi:10.1088/1126-6708/2008/12/015.
  arXiv:0810.1563 [hep-th]

\bibitem{Hartnoll:2008vx}
  S.~A.~Hartnoll, C.~P.~Herzog, G.~T.~Horowitz,
  Phys.\ Rev.\ Lett.\  {\bf 101}, (2008) 031601.
  doi:10.1103/PhysRevLett.101.031601.
  arXiv:0803.3295 [hep-th]

\bibitem{Lin:2014bya}
  K.~Lin, E.~Abdalla, A.~Wang,
  Int.\ J.\ Mod.\ Phys.\ D {\bf 24}, (2015) 0038.
  doi:10.1142/S0218271815500388.
  arXiv:1406.4721 [hep-th]

\bibitem{Abdalla:2013zra}
  E.~Abdalla, J.~de Oliveira, A.~B.~Pavan, C.~E.~Pellicer,
   \textit{Holographic phase transition and conductivity in three dimensional Lifshitz black hole},
  arXiv:1307.1460 [hep-th].

\bibitem{Pan:2009xa}
  Q.~Pan, B.~Wang, E.~Papantonopoulos, J.~Oliveira, A.~B.~Pavan,
  Phys.\ Rev.\ D {\bf 81},(2010) 106007.
  doi:10.1103/PhysRevD.81.106007.
  arXiv:0912.2475 [hep-th]

\bibitem{Abdalla:2010nq}
  E.~Abdalla, C.~E.~Pellicer, J.~de Oliveira, A.~B.~Pavan,
  Phys.\ Rev.\ D {\bf 82}, (2010) 124033.
  doi:10.1103/PhysRevD.82.124033.
  arXiv:1010.2806 [hep-th]

\bibitem{kdv}
  E.~Abdalla, J.~de Oliveira, A.~Lima-Santos, A.~B.~Pavan,
  Phys.\ Lett.\ B {\bf 709}, (2012) 276.
  doi:10.1016/j.physletb.2012.02.026.
  arXiv:1108.6283 [hep-th]

 \bibitem{berjefpell}
  B.~Cuadros-Melgar, J.~de Oliveira, C.~E.~Pellicer,
  Phys.\ Rev.\ D {\bf 85}, (2012) 024014.
  doi:10.1103/PhysRevD.85.024014.
  arXiv:1110.4856 [hep-th]

  \bibitem{owenfideljef}
  E.~Abdalla, O.~P.~F.~Piedra, F.~S.~Nunez, J.~de Oliveira,
  Phys.\ Rev.\ D {\bf 88}, no. 6, (2013) 064035.
  doi:10.1103/PhysRevD.88.064035.
  arXiv:1211.3390 [gr-qc]

\bibitem{kaielciojef}
  K.~Lin, J.~de Oliveira, E.~Abdalla,
  Phys.\ Rev.\ D {\bf 90}, no. 12, (2014) 124071.
  doi:10.1103/PhysRevD.90.124071
  arXiv:1409.4066 [hep-th]

\bibitem{strominger}
  A.~Strominger,
  JHEP {\bf 0110}, (2001) 034.
  doi:10.1088/1126-6708/2001/10/034.
  arXiv:0106113 [hep-th]

\bibitem{Buchbinder:1992rb}
  I.~L.~Buchbinder, S.~D.~Odintsov, I.~L.~Shapiro,
  \textit{Effective action in quantum gravity}, (IOP, Bristol, UK, 1992), 413 p.

\bibitem{gao}
   C. Gao , 
   JCAP \textbf{023}, (2010) 1006.
   doi:10.1088/1475-7516/2010/06/023.
   arXiv:1002.4035

\bibitem{sushkov}
   S.~V.~Sushkov,
  Phys.\ Rev.\ D {\bf 80} (2009) 103505.
  doi:10.1103/PhysRevD.80.103505.
  arXiv:0910.0980 [gr-qc]

\bibitem{capozzielo}
   S. Capozziello, G. Lambiase and H. -J. Schmidt, 
   Annalen Phys. \textbf{9}, (2000) 39.
   arXiv: 9906051 [gr-qc]

\bibitem{capozzielo2}
   S. Capozziello, G. Lambiase, 
   Gen. Rel. Grav. \textbf{31}, (1999) 1005.
   doi:10.1023/A:1026631531309.
   arXiv:9901051 [gr-qc]

\bibitem{amendoa2}
   L. Amendola, 
   Phys. Let. B \textbf{301}, (1993) 175.
   doi:10.1016/0370-2693(93)90685-B.
   arXiv: 9302010 [gr-qc]

\bibitem{holo}
   S. Chen, Q. Pan, J. Jing, 
   Chin. Phys. B \textbf{21}, (2012) 040403.
   doi:10.1088/1674-1056/21/4/040403.
   arXiv:1012.3820 [gr-qc]

\bibitem{amendoa1}
   L. Amendola, M. Litterio, F. Occhionero, 
   Int. J. Mod. Phys. A \textbf{05}, 3861 (1990).
   doi:10.1142/S0217751X90001653.

\bibitem{quasi1}
   R. A. Konoplya, Z. Stuchlik, A. Zhidenko, \textit{A massive non-minimally coupled scalar field in Reissner-Nordstr\"om spacetime: long-lived quasinormal modes and instability}, arXiv:1808.03346v1.

\bibitem{quasi2}
   S. Yu, C. Gao, \textit{Quasinormal modes of static and spherically symmetric black holes with the derivative coupling}, arXiv:1807.05024v1.

\bibitem{quasi3}
   S. Chen, J. Jing,
  Phys. Rev. D {\bf 90}, (2014) 124059.
  doi:10.1103/PhysRevD.90.124059.
  arXiv:1408.5324 [gr-qc].

\bibitem{esta}
   R. A. Konoplya, A. Zhidenko,
   Phys. Rev. Lett. \textbf{103}, (2009) 161101, doi:10.1103/PhysRevLett. 103.161101, arXiv:0809.2822 [hep-th]; R. A. Konoplya, A. Zhidenko, 
Phys. Rev. D \textbf{78}, (2008) 104017, doi:10.1103/PhysRevD.78.104017, arXiv:0809.2048 [hep-th]; R. A. Konoplya and A. Zhidenko, 
   Phys. Rev. D \textbf{77}, (2008) 104004, doi:10.1103/PhysRevD.77.104004,
   arXiv:0802.0267 [hep-th]; A. Zhidenko, 
Class. Quant. Grav. \textbf{21}, (2004) 273, arXiv:0307012 [gr-qc];  R. A. Konoplya, A. Zhidenko, 
 JHEP {\bf 0406},(2004) 037, doi:10.1088/1126-6708/2004/06/037, arXiv:0402080 [hep-th/];


\bibitem{Chen} S. Chen, J. Jing, 
   Phys. Lett. B \textbf{691}, (2010) 254, doi:10.1016/j.physletb.2010.06.041,
   arXiv:1005.5601 [gr-qc].
               S. Chen, J. Jing, \textit{Dynamical evolution of a scalar field coupling to Einstein's tensor in the Reissner-Nordstr\"om black hole spacetime},
 arXiv:1007.2019.

\bibitem{Molina} C. Molina, D. Giugno, E. Abdalla, A. Saa, 
 Phys. Rev. D \textbf{69}, (2004) 104013.
 doi:10.1103/PhysRevD.69.104013.
 arXiv:0309079 [gr-qc]



\bibitem{Bin} Da-Ping Du, Bin Wang, Ru-Keng Su, 
Phys. Rev. D {\bf 70}, (2004) 064024.
doi:10.1103/PhysRevD.70.064024.
arXiv:0404047[hep-th]

\bibitem{slovaka} Z.~Stuchlik, S.~Hledik, 
 Acta physica slovaca {\bf 52} no. 5.
 arXiv:0803.2685 [gr-qc].

\bibitem{rosa}
    R. A. Konoplya, A. Zhidenko, 
Rev. Mod. Phys., \textbf{83}, (2011) 793.
doi:10.1103/RevModPhys.83.793.
arXiv:1102.4014 [gr-qc]

\bibitem{Cardoso}
    V. Cardoso, J. O. S. Lemos, 
Phys. Rev. D \textbf{67},(2003) 084020.
doi:10.1103/PhysRevD.67.084020.
arXiv:0301078 [gr-qc]

\bibitem{MolinaII}
   C. Molina, 
Phys. Rev. D \textbf{68}, (2003) 064007.
doi:10.1103/PhysRevD.68.064007.
arXiv:0304053 [gr-qc]

\bibitem{dsneg}
    P. R. Brady, C. M. Chambers, W. Krivan, P. Laguna,
Phys. Rev. D \textbf{55}, (1997) 7538.
arXiv:9611056 [gr-qc]


\end{thebibliography}
\end{document}